\documentclass[prd,showpacs,nofootinbib,amsmath,amssymb,superscriptaddress,twocolumn]{revtex4-1}

\usepackage{dcolumn}
\usepackage{bm}
\usepackage{amsmath,amssymb,bbm,mathrsfs,nicefrac}
 \usepackage{array,multirow}
 \usepackage{float}
\usepackage{tikz}
\usepackage{amsmath}
\usepackage{amssymb}
\usepackage{graphicx,textcomp,float,gensymb,wrapfig, enumitem,comment,dsfont,framed,slashed,appendix,wrapfig,xcolor}
\usepackage{ bbold }
\usepackage[normalem]{ulem}
\usepackage{braket} 
\usepackage{ mathrsfs }
\usepackage{etoolbox,hyperref,makecell}
\usepackage{feynmp}
\usetikzlibrary{arrows.meta}
\usepackage{empheq}
\usepackage[normalem]{ulem}
\usepackage{scrextend}
\usepackage{slashed}
\newcommand{\bea}{\begin{eqnarray}}  
\newcommand{\eea}{\end{eqnarray}}

\newcommand{\1}{{\bf 1}}
\definecolor{darkgreen}{rgb}{0.2, 0.3, 0.1}
\definecolor{orchid}{rgb}{127, 0, 255}

\usepackage[utf8]{inputenc}

\begin{document}

\title{Validity of a finite temperature expansion for dense nuclear matter}

\author{Debora Mroczek}
\author{Nanxi Yao}
\author{Katherine Zine}
\author{Jacquelyn Noronha-Hostler}
\affiliation{Illinois Center for Advanced Studies of the Universe, Department of Physics, University of Illinois at Urbana-Champaign, Urbana, IL 61801, USA}
\author{Liam Brodie}
\affiliation{Department of Physics, Washington University in St.~Louis, 63130 Saint Louis, MO, USA}
\author{Veronica Dexheimer}
\affiliation{Center for Nuclear Research, Department of Physics, Kent State University, Kent, OH 44243 USA}
\author{Alexander Haber}
\affiliation{Mathematical Sciences and STAG Research Centre, University of Southampton, Southampton SO17 1BJ, United Kingdom}
\affiliation{Department of Physics, Washington University in St.~Louis, 63130 Saint Louis, MO, USA}
\author{Elias R. Most}
\affiliation{TAPIR, Mailcode 350-17, California Institute of Technology, Pasadena, CA 91125, USA}
\affiliation{Walter Burke Institute for Theoretical Physics, California Institute of Technology, Pasadena, CA 91125, USA}

\date{\today}

\begin{abstract}
In this work we provide a new, well-controlled expansion of the equation of state of dense matter from zero to finite temperatures ($T$) while covering a wide range of charge fractions ($Y_Q$), from pure neutron to isospin symmetric nuclear matter. 
Our expansion can be used to describe neutron star mergers using the equation of state inferred from neutron star observations. 
We discuss how knowledge from low-energy nuclear experiments and heavy-ion collisions can be directly incorporated into the expansion. 
We also suggest new thermodynamic quantities of interest that can be calculated from theoretical models or directly inferred by experimental data that can be used to infer the finite temperature equation of state.
With our new method, we can quantify the uncertainty in our finite $T$ and $Y_Q$ expansions without making assumptions about the underlying  degrees of freedom. 
We can reproduce results from a microscopic equation of state up to $T=100$ MeV for baryon chemical potential $\mu_B\gtrsim 1100$ MeV ($\sim1-2 \ n_{\rm sat}$) within $5\%$ error, with even better results for larger $\mu_B$ and/or lower $T$. 
We investigate the sources of numerical and theoretical uncertainty and discuss future directions of study.

\end{abstract}

\maketitle

\section{Introduction}
    Quantum chromodynamics (QCD) is a non-perturbative theory in the ranges of temperature ($T$), baryon number density ($n_B$), and charge fraction ($Y_Q = n_Q/n_B$, where $Q$ denotes electric charge) relevant for astrophysical phenomena such as neutron star formation in supernovae and neutron star binary mergers. 
    This feature of QCD poses a modeling challenge since the equilibrium, i.e.~the equation of state (EoS), and transport properties of nuclear matter relevant for neutron star formation and mergers cannot be determined from first-principle methods. 

    The nuclear matter created in these events also spans a wide range of the QCD phase diagram, requiring knowledge of equilibrium and transport properties of QCD for $T \lesssim 100$ MeV, $ 0 \leq n_B \lesssim \10$ $n_{\rm sat}$, where $n_{\rm sat} \equiv 0.16$ fm$^{-3}$ is the nuclear saturation density, and $0.01 \lesssim Y_Q \lesssim 0.5 $ \cite{Most:2018eaw,Most:2019onn}. 
    At $T=0$, these ranges in $n_B$ and $Y_Q$ cover different phases of matter -- from nuclei to a neutron-proton gas, and likely undiscovered exotic phases at higher densities. 
    The correct description at $T>0$ is even more uncertain. 

    A unified theoretical model of QCD over such wide ranges in phase space is not currently available. Many EoS based on parametrizations of nuclear forces at densities above $n_{\rm sat}$ have been calculated at zero and finite temperatures \cite{Lattimer:1991nc,Shen:1998gq,Hempel:2009mc,Steiner:2012rk,Shen:2011fc,Shen:2011kr,Alford:2022bpp,Dexheimer:2008ax,Dexheimer:2009hi, Papazoglou:1998vr, Steinheimer:2010ib,Motornenko:2019arp,Gulminelli:2015csa,Raduta:2018aqy,Typel:2018wmm, Togashi:2017mjp,Fujimoto:2021dvn}, but these frameworks fix the degrees of freedom and interactions in the system, and thus, span a limited range of physics. Alternative schemes have been introduced which rely on parametrizations of the thermal component of the EoS, allowing for zero-temperature models to be extended to finite temperatures. A method known as the ``hybrid" approach is a widely used parametrization of the thermal EoS which assumes thermal contributions can be modeled as an ideal fluid and added to any cold EoS \cite{1993A&A...268..360J}. Though popular in the literature (for a review see, e.~g., Ref.~\cite{Baiotti:2016qnr}), Ref.~\cite{Raithel:2019gws} demonstrated that the ideal-fluid approximation may not be valid in the temperature and density regimes relevant for astrophysical phenomena.
    
    A new approach was introduced in Ref.~\cite{Raithel:2019gws} based on an effective mass parametrization of the thermal contributions to the EoS across different density regimes, assuming the degrees of freedom to be protons, neutrons, and electrons (npe). While npe matter describes a large class of cold, beta-equilibrated matter models in the literature, many models predict additional degrees of freedom like strange baryons, baryon resonances, and quarks (see, e.g., Refs.~\cite{Dexheimer:2008ax,Dexheimer:2009hi}). The class of EoS containing degrees of freedom beyond npe cannot be described by the procedure introduced in Ref.~\cite{Raithel:2019gws}. 
    
    In this work, we present a framework which allows for the extension of any cold, beta-equilibrated EoS without making assumptions about the underlying degrees of freedom. This extension requires three steps. 
First, the $T=0$ EoS must be extended beyond beta-equilibrium to arbitrary values of $Y_Q$ between 0, corresponding to pure neutron matter, and 0.5, which corresponds to symmetric nuclear matter. 
For this step we rely on the fact that any $T=0$, beta-equilibrated EoS can be expanded across a range of $Y_Q$ via a parametrization of the symmetry energy \cite{Yao:2023yda}. 
Second, we need to account for finite temperature effects. 
We start with symmetric nuclear matter and perform a Taylor expansion in $(T/\mu_B)$. 
We motivate this expansion with pedagogical examples of fermionic and conformal matter and show that an expansion in $(T/\mu_B)$ fully captures the temperature dependence of the EoS under reasonable assumptions. 
Last, we argue that the charge fraction dependence of finite temperature effects can be modeled with an expansion around symmetric nuclear matter in the isospin asymmetry parameter $\delta=1-2Y_Q$. The result is an EoS that spans arbitrary values of $Y_Q$ and finite $T$ which can be implemented in simulations of neutron star mergers and heavy-ion collisions. 

Our framework depends on the value of the constants of the symmetry energy expansion for which some experimental and theoretical guidance is available and which have been extensively discussed in the literature\footnote{We refer the reader to Ref.~\cite{Yao:2023yda} and references therein for a discussion of these constants, available experimental constraints, and possible additional constraints from thermodynamics.}. Additionally, two density-dependent coefficients are introduced to account for finite temperature effects. We argue that these functions can be studied from microscopic models or extracted directly from heavy-ion collision data \cite{Nana:2024okk}. 

We test the error introduced by the two new expansions against a relativistic mean-field EoS and find that there are a few theoretical and numerical challenges to consider when performing an expansion of the $T=0$ EoS to finite $T$. We breakdown these sources of error and uncertainty and argue that they can be mitigated with improved theoretical calculations and future experimental data. Though some numerical error is inevitable, even with perfect knowledge of the expansion coefficients, we show that this error is small in the region of the phase diagram relevant for neutron star mergers. 

We present here the component of the theoretical framework which accounts for finite temperature effects. We do not include a review of the expansion of the symmetry energy at $T=0$ (we refer the reader to Ref.~\cite{Yao:2023yda}) for brevity and to focus on novel contributions that are of broader interest. A separate manuscript discussing the results of the implementation of our method in realistic simulations of neutron star mergers is currently in preparation \cite{numericalimplementation}.

We use natural units $\hbar=c=1$, unless otherwise specified. Repeated indices imply summation (Einstein summation convention), unless otherwise specified. We assume that the nuclear saturation density is $n_{\rm sat}$ is 0.16 fm$^{-3}$. In this study, we neglect leptons and consider only the QCD EoS. Thus, $Y_Q$ refers to the fraction of charge baryons (hadrons or quarks) in the system.

\section{Finite temperature effects}
\subsection{Formalism}
We begin by expanding the EoS from zero to finite temperature without taking into account charge fraction effects. If we assume that we know the pressure ($p$) at $T=0$ for all values of $\vec{\mu}$, where $\vec{\mu}$ is the set of chemical potentials associated with conserved charges of the system.
In this work we assume baryon number and electric charge such that $\vec{\mu}=\left\{\mu_B,\mu_Q\right\}$.
That is, we know $p(T=0,\vec{\mu})$. We can write $p(T,\vec{\mu})$ as
\begin{align}\label{eq:exp_0}
    p(T,\vec{\mu}) &= p(T=0,\vec{\mu}) +\left(\dfrac{d p}{d T}\right)_{\vec{\mu},T=0}T\nonumber\\
    & + \dfrac{1}{2}\left(\dfrac{d^2 p}{d T^2}\right)_{\vec{\mu},T=0}T^2 + \dfrac{1}{2}\left(\dfrac{d^3 p}{d T^3}\right)_{\vec{\mu},T=0}T^3\dots
\end{align}
Recall that the third law of thermodynamics requires that the entropy ($s = dp/dT|_{\vec{\mu}}$) smoothly approaches a constant value as the temperature approaches zero. Here, we assume that this value is zero, such that the linear term in the above expression vanishes\footnote{This approximation is valid for the types of systems discussed here, which include ideal, weakly-interacting, and mean-field treatments of fermions. However, nonzero entropy at zero $T$ can arise in a system which contains multiple ground states or residual entropy from being ``frozen" in a state which does not minimize the energy (e.~g.~an imperfect crystal).}.

We can rewrite Eq.~\ref{eq:exp_0} in terms of the pressure at zero temperature and derivatives of the entropy,
\begin{align}\label{eq:exp_1}
    p(T,\vec{\mu}) &= p(T=0,\vec{\mu}) + \dfrac{1}{2}\left(\dfrac{d s}{d T}\right)_{\vec{\mu}, T=0}T^2 \nonumber\\
    & + \dfrac{1}{6}\left(\dfrac{d^2 s}{d T^2}\right)_{\vec{\mu}, T=0}T^3\dots
\end{align}

This expansion requires knowledge of the pressure at zero temperature, $p(T=0,\vec{\mu})$, and the coefficients, $\left(d^n s/d T^s\right)_{\vec{\mu}, T=0}$. 
We choose to write this as a dimensionful expansion i.e. in terms of power of $T$.
However, given that the expansion is calculated at fixed $\vec{\mu}$ then this is equivalent to a $T/\mu$ expansion.

\subsection{Expansion parameter and additional assumptions}\label{subsec:exp_parameter}
The expansion parameter is $(T/\mu)$, where $\mu$ is the total chemical potential\footnote{Since each $\mu_i$ is associated with a conserved charge that is held constant for derivatives with respect to $T$, in Eqs.~(\ref{eq:exp_0}) and (\ref{eq:exp_1}) we assume the $\vec{\mu}$ dependence is absorbed in the expansion coefficients which have dimensions of MeV$^{4-n}$.}. 
As a first step, we set $\mu_Q = 0$. We will later account for nonzero $\mu_Q$ when we discuss the inclusion of isospin asymmetry effects. 
We assume strangeness is in chemical equilibrium, i.e., $\mu_S = 0$. 
We make this choice because the relevant time scale for neutron star mergers is ms, which allows for strangeness to chemically equilibrate, such that $\mu_S=0$ throughout the merger. 
That is not the case for, e.~g.~heavy-ion collisions, where the evolution scale is a few fm/c ($\sim10^{-24}$ s). 
Furthermore, in heavy-ion collisions, net-strangeness is conserved such that one assumes (global) strangeness neutrality. 
However, large local fluctuations of strangeness charge density can occur, leading to local fluctuations of $\mu_S$ \cite{Plumberg:2024leb}. 

\begin{figure}
    \centering
    \includegraphics[width=\linewidth]{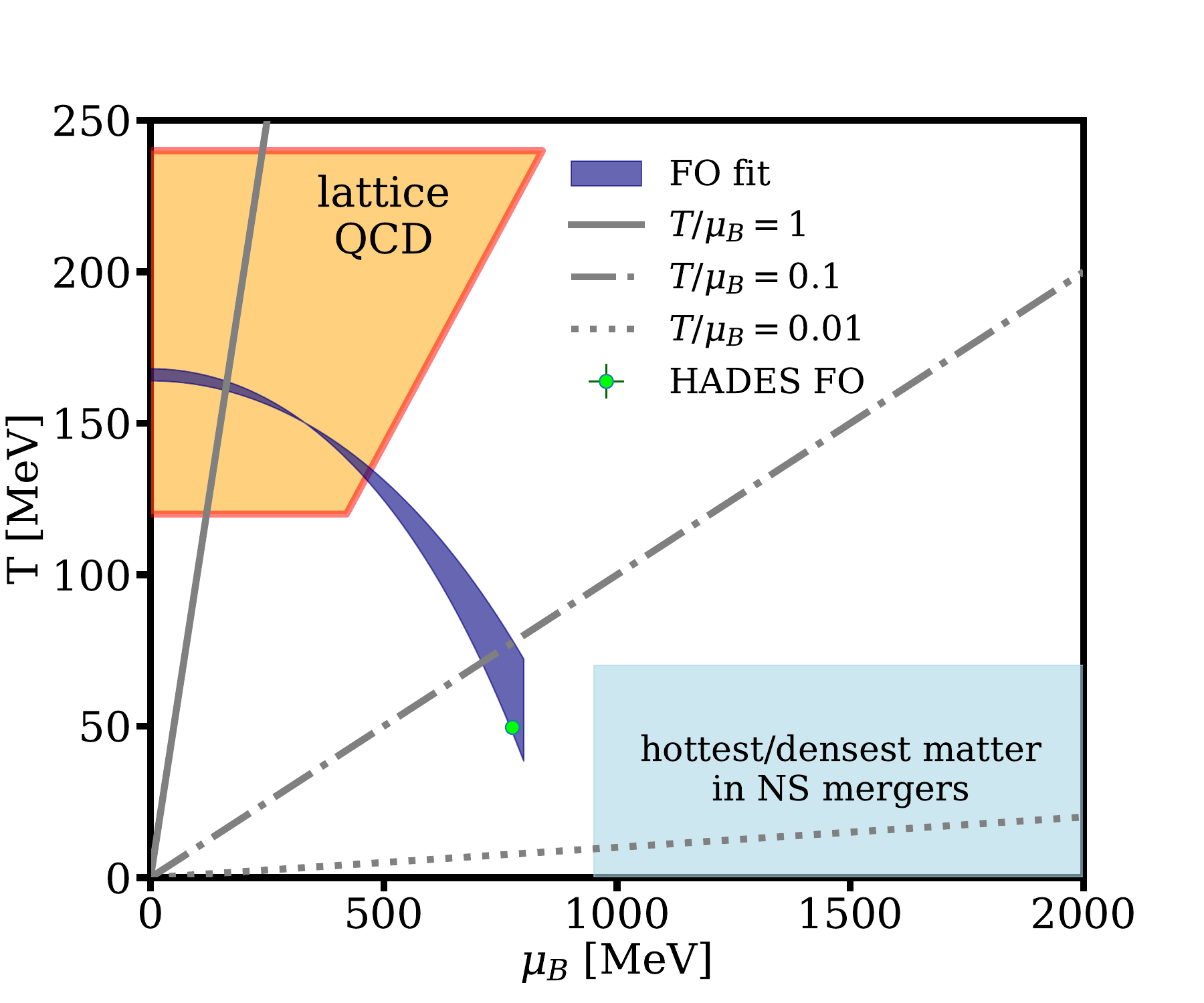}
    \caption{Comparing the magnitude of the finite temperature expansion parameter $(T/\mu_B)$ to the region where $(\mu_B)/T \leq 3.5$ the highest value where uncertainty quantification from lattice QCD is available for the EoS \cite{Parotto:2024sbw}. Also shown are the estimated freeze-out line from thermal fits of particle yields from several heavy-ion collision experiments at different center-of-mass energies \cite{Cleymans:2005xv} and the estimated freeze-out point from thermal fits of particle yields at $\sqrt{s_{\rm NN}} = 2.4$ GeV at the HADES experiment \cite{Harabasz:2020sei,Motornenko:2021nds}. The high-density region of the phase diagram probed by neutron star mergers shown in light blue ($T\lesssim 70$ MeV, $\mu_B\gtrsim 950$ MeV \cite{Most:2018eaw,Most:2019onn}) is well within the domain where the expansion parameter is small, $(T/\mu_B) < 0.1$. Heavy-ion collisions in the few-GeV range would also be in the regime where $(T/\mu_B)$
is small, though one should keep in mind the liquid-gas phase transition and a possible high-temperature critical point/first-order phase transition associated with deconfinement. }
    \label{fig:phasediagram}
\end{figure}

Another strong motivation for making assumptions based on the relevant scales for neutron star mergers is that the expansion parameter $(T/\mu)$ is small in the regime relevant for mergers. 
In Fig.~\ref{fig:phasediagram}, we show the $T-\mu_B$ plane and mark in gray the regions corresponding to $(T/\mu_B)$ values smaller than 0.01 (dotted), 0.1 (dot-dashed), and 1 (solid). 
For reference, we include the region above $T=110$ MeV and $(\mu_B/T) \leq 3.5$, where uncertainty estimation from lattice QCD results \cite{Parotto:2024sbw} is available, the estimated freeze-out (FO) line from a number of heavy-ion experiments \cite{Cleymans:2005xv}, and the estimated FO point at center-of-mass-energy, $\sqrt{s_{\rm NN}} = 2.4$ GeV for the High Acceptance Di-Electron Spectrometer (HADES) experiment \cite{Harabasz:2020sei,Motornenko:2021nds}. 
From the FO fit, we see that intermediate to low energy heavy-ion collisions, span the domain where $0.1 < (T/\mu_B) < 1$, except the few-GeV region explored by HADES and the fixed target program at the Relativistic Heavy-Ion Collider (RHIC) \cite{STAR:2017sal}, where $ (T/\mu_B) < 0.1$.
On the other hand, mergers of neutron stars are estimated to probe the phase diagram below 70 MeV in temperature and above 950 MeV in $\mu_B$ \cite{Most:2018eaw,Most:2019onn}, placing these events well-within the $(T/\mu_B) < 0.1$ region. We show the region corresponding to the hottest and densest matter created in neutron star mergers in light blue. 
Therefore, the expansion is more suitable for describing the nuclear matter created in neutron star mergers. 
Nonetheless, it is possible to gain information about the expansion coefficients from heavy-ion collisions, as we will discuss later in Sec.~\ref{sec:yq_effects}. 

A significant complication arising from this choice is the scaling of thermodynamic quantities near the liquid-gas phase transition. 
For symmetric nuclear matter, the transition is estimated to happen around $\mu_B=$900 MeV with a critical point at $T\sim 15-20$ MeV (see, e.~g., Refs.~\cite{Vovchenko:2020lju,Elliott:2012nr,Karnaukhov:2008be}). Thus, our expansion is only applicable above the transition density, and not able to capture the critical scaling at finite temperatures arising from the presence of a critical point. However, there are frameworks to map a critical point onto a smooth EoS \cite{Parotto:2018pwx,Kapusta:2022pny}. 
Furthermore, there is significant community effort to combine EoS which have limited regimes of applicability into families of unified, thermodynamically consistent EoS \cite{MUSES:2023hyz,ReinkePelicer:2025vuh}. 
Therefore, despite the limited applicability of our expansion, it still allows for any zero-temperature, high-density EoS to be extended to finite temperatures, and solutions already exist for the limitations we identified above. 

Lastly, note that at low $\mu_B$ and high T, a $(\mu_B/T)$ expansion has been the standard in the literature for many years (see, e.~g., Ref.~\cite{Parotto:2018pwx}), though the expansion parameter can be as large as $(\mu_B/T) = $3.5. In contrast, the regime we propose for a $(T/\mu_B)$ expansion would require at most $(T/\mu_B) \sim 0.1$. 

\subsection{Physical motivation}
Albeit simple, the expansion defined in Eq.~\ref{eq:exp_1} captures the behavior of many systems relevant for neutron star mergers and low-energy heavy-ion collisions. We briefly discuss two large classes of EoS that can be expressed as a power series in $T$. 

\subsubsection{Example 1: Fermi gases}

It is well-known that for an ideal Fermi gas at $T \ll T_F$, where $T_F$ is the Fermi temperature\footnote{For an ideal gas, $T_F=\mu$.}, the pressure scales with $T^2$ at leading order in $T$.
That is because when $T \ll T_F$, the integral which appears in the calculation of thermodynamic quantities can be written as
\begin{align}\label{eq:Sommerfeld}
    I & = \int_0^\infty d\varepsilon_j f(\varepsilon_j) \phi(\varepsilon_j)  = \int_0^{\mu} d\varepsilon_j\phi(\varepsilon_j) \nonumber\\
    & + \dfrac{\pi^2}{6} T^2\left(\dfrac{d\phi}{d \varepsilon_j}\right)_{\varepsilon_j = \mu}  + \mathcal{O}(T^{2n}),
\end{align}
where $\varepsilon_j$ is the energy of the $j^{\rm th}$ state of the system, $f(\varepsilon_j)$ is the Fermi distribution, $\phi(\varepsilon_j) = A \varepsilon_j^\alpha$ for a constant $A$ and $\alpha \geq 1/2$, $\mu$ is the chemical potential, and $n > 1$ is an integer.
This procedure is known as the Sommerfeld expansion. 

The expression for the pressure is given by $\phi(\varepsilon_j) = \frac{4}{3}C\varepsilon_j^{1/2}$, where $C=m^{3/2}/\sqrt{2}\pi^2$ is a constant, such that applying the Sommerfeld expansion to $\mathcal{O}(T^2)$ yields,

\begin{align}
    p(T,\mu) = \dfrac{8}{15}C\mu^{5/2} + \dfrac{\pi^2}{3}C\mu^{1/2}T^2,
\end{align}
which is in the form given by Eq.~\ref{eq:exp_1}. Thus, an ideal Fermi gas in the regime where $T \ll T_F$ can be described by our finite temperature expansion. 

\emph{Beyond the ideal Fermi gas:}
If a system has interactions which shift the energy levels but do not cause them to rearrange, it can be approximated by an ideal gas for which the masses are screened by the interactions \cite{Landau:1956zuh,Baym:1975va}. 
This behavior is captured by an effective mass term $m_*$ and an energy shift $U$, such that $T_F = E_F - m_* = \sqrt{k_F^2 + m_*^2} + U$ (see Ref.~\cite{Cruz-Camacho:2024odu} for an example). 
Mean-field treatments of nuclear matter where interactions are mediated via mesons can be described by this formalism. Our expansion is especially suited for approximating these systems, also known as Fermi liquids. 

Note, however, that the notion of a Fermi temperature is not well-defined for a system which contains different particles, so we cannot draw a direct parallel between the Fermi temperature for a given particle at given $\mu$ and the applicability of our expansion. Nonetheless, those values can be used as estimates for the breakdown of the expansion, as the behavior of individual particles can no longer be estimated by a Sommerfeld approximation. 

\subsubsection{Example 2: a class of conformal EoS}
The most general form of a conformal EoS is $p = T^4f(\mu/T)$, where $f$ is an arbitrary function which must be even in $\mu$ and such that the resulting $p$ respects thermodynamic laws and stability/causality relations. Consider a class of conformal EoS which can be expanded as a power series in $T$, such that the dimensionless pressure is 
\begin{align}
    \frac{p}{\mu^4} = c + \frac{b}{\mu^2}T^2+\frac{a}{\mu^4}T^4,
\end{align}

where $a,b$ and $c$ are constants. 

An example of EoS that has the form above is a massless quark-gluon plasma, for which $a \approx 4.63 $, $b \approx 0.14$, $c\approx0.0094$  if the number of colors is $N_c=3$ and the number of flavors is $N_f=2.5$ \cite{Chaudhuri:2012yt}. 

\emph{Smoothly connecting the EoS across densities:}  At high-densities, beyond those realized in neutron stars, QCD is expected to become perturbative and approximately conformal \cite{Kurkela:2016was,Gorda:2023mkk}. Within our framework, this regime can be connected to lower-densities via the expansion coefficients as long as the transition from hadrons to quarks is a crossover.

\section{Charge fraction dependence of finite temperature effects}\label{sec:yq_effects}

\subsection{Formalism}
Now that we have motivated a finite temperature expansion for symmetric nuclear matter, we want to parameterize the charge fraction dependence of the EoS at finite temperatures, so that we can obtain $p(\vec{\mu}, T)$. 
Recall that at $\mathcal{O}(T^2)$, there are two functions of $\mu_B$ that need to be specified, the zero-temperature pressure, $p(T=0,\mu_B)$, and the expansion coefficient, $(d s/d T)_{\mu_Q, T=0}$. 
The symmetry energy expansion \cite{Bombaci:1991zz} is the standard procedure adopted by the community to expand an EoS at a fixed charge fraction (typically symmetric nuclear matter for which $Y_Q = 0.5$) to arbitrary $Y_Q$. 
The key assumption for this expansion is that the EoS is isospin symmetric and therefore only even terms contribute to the expansion around $Y_Q = 0.5$. Here, we will adopt the same assumption, and define a similar expansion for the $T^2$ coefficient in Eq.~\ref{eq:exp_1}.

Before we continue, let us define acronyms for two limits of QCD which are useful for the derivation. We denote as symmetric nuclear matter (SNM) systems which contain an equal number of protons and neutrons, i.e.~$Y_Q = 0.5$. Another useful limit is pure neutron matter (PNM) which describes systems made up entirely of neutrons ($Y_Q=0$). From these two limits, we can define a parameter which captures the isospin asymmetry of the system, 
\begin{equation}\label{eqn:asym_para}
    \delta=1-2 Y_Q,
\end{equation}
and quantify the deviation from symmetric nuclear matter. That is $0 \leq \delta \leq 1$, where $\delta = 0$ is SNM and $\delta = 1$ is PNM\footnote{This assumes neutron and protons are the dominant particles. Any particles with negative electric charge (e.g. $\Sigma^-$) or double electric charge $\Delta^{++}$ are sub-leading effects.}. Note that this definition implies the strangeness charge fraction is zero. That is, we do not allow for strange degrees of freedom. 
Then, the symmetry energy ($E_{\rm sym}$) is the difference between these two limits. If we only consider the terms up to $\delta^2$, we have,
\begin{equation}
    E_{\rm{sym},2}=\frac{E_{\rm PNM}-E_{\rm SNM}}{N_B \left(\delta_{\rm PNM}-\delta_{\rm SNM}\right)^2}=\dfrac{E_{\rm PNM}-E_{\rm SNM}}{N_B},
\end{equation}
where $N_B$ is the total number of baryons in the system.
The, the energy per baryon for asymmetric nuclear matter (ANM), that is, $0 < \delta < 1$, can be written as
\begin{equation}
    \dfrac{E_{\rm ANM}}{N_B} = \dfrac{E_{\rm SNM}}{N_B} +  E_{\rm sym,2}\delta^2 + \mathcal{O}(\delta^4).
\end{equation}

In our approach, because we are interested in finite temperature effects, we look at the symmetry \emph{specific entropy}, which we define as $\tilde{S} = s/n_B$, where $s$ and $n_B$ are the entropy and baryon number \emph{densities}. Then, the $\delta^2$ symmetry specific entropy term is,
\begin{eqnarray}
    \tilde{S}_{\rm sym,2}&=&\frac{s_{\rm PNM}/n_B-s_{\rm SNM}/n_B}{\left(\delta_{\rm PNM}-\delta_{\rm SNM}\right)^2}\nonumber\\
    &=&\frac{s_{\rm PNM}-s_{\rm SNM}}{n_B}.
\end{eqnarray}

We can expand $ \tilde{S}~(T,n_B,Y_Q)$ around SNM just as it is typically done for the symmetry energy expansion, assuming the function is even around the expansion point. The expansion is then, 
\begin{equation}\label{eqn:sexpan}
     \tilde{S}(T,n_B,\delta)=\frac{s_{\rm{SNM}}}{n_B}(T,n_B)+ \frac{1}{2}\tilde{S}_{\rm{sym},2}(T,n_B)\delta^2+\mathcal{O}\left(\delta_{\rm }^4\right) ,
\end{equation}
and, consequently, 
\begin{equation}
    \tilde{S}_{\rm{sym},2}(T,n_B)\equiv \frac{\partial^2 \tilde{S}(T,n_B,\delta)}{\partial \delta^2}\Bigg|_{T, n_B, \delta=0} ,
\end{equation}

From this point on we drop higher-order terms and rewrite Eq.\ (\ref{eqn:sexpan}) as
\begin{align}\label{eqn:snB_finiteT}
     \tilde{S}(T,n_B,\delta)=\frac{s_{\rm{SNM}}}{n_B}(T,n_B)+ \frac{1}{2}\tilde{S}_{\rm{sym},2}(T,n_B)\delta^2.
\end{align}

Now, we can take the temperature derivative, 
\begin{widetext}
\begin{equation}\label{eqn:dsdT_YQexpan}
    \frac{\partial \tilde{S}(T,n_B,\delta)}{\partial T}\Bigg|_{T=0} =\frac{1}{n_{B}}\frac{\partial s_{\rm{SNM}}(T,n_B)}{\partial T}\Bigg|_{T=0}+\frac{1}{2}\frac{\partial^3\tilde{S}(T,n_B,\delta)}{\partial T\partial \delta^2}\Bigg|_{T=\delta=0} \delta^2,
\end{equation}
\end{widetext}
where we can factor out $n_B$ and $\delta$ terms because they are constant.

Using Eq.\ (\ref{eqn:dsdT_YQexpan}), if we know the first term, 
\begin{align*}
    \frac{\partial s_{\rm{SNM}}(T,n_B,Y_Q)}{\partial T}\Big|_{T=\delta=0}
\end{align*} 
and the second term
\begin{align*}
    \frac{\partial^3\tilde{S}(T,n_B)}{\partial T\partial \delta^2}\Big|_{T=\delta=0},
\end{align*}
we can calculate 
$\frac{\partial s}{\partial T}\big|_{T=0}$ along any slice of constant $Y_Q$, which is the coefficient needed for the finite $T$ expansion of the pressure.

Since we require the information about $\frac{\partial s}{\partial T}\big|_{T=0,\vec{\mu}}$ 
along a grid of fixed $\vec{\mu}$ in
Eq.\ (\ref{eq:exp_1}), but the expansion that we have derived above is in terms of fixed 
$n_B,\delta$ (or rather $n_B,Y_Q$ since $\delta$ and $Y_Q$ have a direct relationship), i.e., 
$\frac{\partial s}{\partial T}\big|_{T=0,n_B,Y_Q}$, we must perform a mapping between the two quantities, i.e., 
\begin{equation}
    \frac{\partial s}{\partial T}\Big|_{T=0,\vec{\mu}} \underset{?}{\leftrightarrow} \frac{\partial s}{\partial T}\Big|_{T=0,n_B,Y_Q} ,
\end{equation}
The Jacobian transformation between the thermodynamical basis $(\vec{\mu},T)$ and $(T,n_B,Y_Q)$ yields 
\begin{align}
    \frac{\partial s}{\partial T}\Big|_{T=0,\vec{\mu}} &= \dfrac{\partial s}{\partial T}\Big | _{n_B,Y_Q,T=0} + \\
    &c_1\dfrac{\partial s}{\partial n_B}\Big| _{T=0,Y_Q} + c_2\dfrac{\partial s}{\partial Y_Q}\Big | _{n_B,T=0}.
\end{align}

In Appendix \ref{app:Jacobian}, we show explicitly the conversion between these two thermodynamic ensembles and demonstrate that $c_1$ and $c_2$, namely, the correction terms, go to zero at $T=0$.  
That is, 
\begin{equation}
    \frac{\partial s}{\partial T}\Big|_{T=0,\vec{\mu}} \sim \frac{\partial s}{\partial T}\Big|_{T=0,n_B,Y_Q}
\end{equation}
where this also implies the equivalence between
\begin{equation}
    \frac{\partial s}{\partial T}\Big|_{T=0,\vec{\mu}} \sim \frac{\partial s}{\partial T}\Big|_{T=0,n_B,\delta}.
\end{equation}

Finally, we can write a finite temperature expansion of SNM to arbitrary values of $Y_Q$ on a grid of $(T,\vec{\mu})$ to $\mathcal{O}(T^2,\delta^2)$ as

\begin{widetext}
\begin{equation}\label{eqn:final_dsdT_YQexpan}
     p(T,\vec{\mu}) = p_{T=0}(\vec{\mu}) + \frac{1}{2}\left[\frac{\partial s_{\rm{SNM}}(T,n_B)}{\partial T}\Bigg|_{T=0}
     +\frac{n_{B}(T=0,\vec{\mu})}{2} \frac{\partial^3\tilde{S}(T,n_B,\delta)}{\partial T\partial \delta^2}\Bigg|_{T=\delta=0}\delta^2  \right]T^2,
\end{equation}
\end{widetext}
where the dependence of $n_B$ and $\delta$ on $\vec{\mu}$ for the derivative terms is implied.   

\subsection{Connection to heavy-ion collisions}
It may be possible to use heavy-ion collision (HIC) data to constrain the terms in Eq.~\ref{eqn:final_dsdT_YQexpan}, namely, $\frac{\partial s_{\rm{SNM}}(T,n_B,\delta=0)}{\partial T}$ and $\frac{\partial^3\tilde{S}_{\rm{SNM},2}(T,n_B)}{\partial T\partial \delta^2}\big|_{\delta=0}$. 
At chemical freeze-out, a model for the EoS can be used to extract the temperature and chemical potentials from identified particles yields, ratios, and fluctuations.  
Commonly, it is assumed that at freeze-out the system can be described by an ideal hadron resonance gas model \cite{Alba:2020jir}.  
However, more realistic models can be used instead, such as a van der Waals EOS \cite{Poberezhnyuk:2019pxs}, lattice QCD \cite{Borsanyi:2013hza,Borsanyi:2014ewa}, or other effective models. 
In fact, any framework that provides microscopic information about the densities of specific particle species could then be used to extract these temperatures and chemical potentials from experimental data given a specific beam energy $\sqrt{s_{NN}}$ and a colliding species that fixes $Y_Q^{\rm{HIC}}=Z/A$.  

Regardless of the underlying model, the general procedure is the same. The EOS must be 4D in terms of $T,\mu_B,\mu_S,\mu_Q$ wherein $\mu_S$ and $\mu_Q$ are constrained by
\begin{eqnarray}
    \langle n_S\rangle &=& 0 ,\label{eqn:str_neu}\\
    \langle n_Q\rangle &=& Y_Q^{\rm{HIC}}\langle n_B\rangle , \label{eqn:charge}
\end{eqnarray}
where the brackets denote net charge densities,
such that we can always determine 
\begin{equation}\label{eqn:chempots}
    \mu_S\left(T,\mu_B\right),\quad \mu_Q\left(T,\mu_B\right) ,
\end{equation}
from $T$ and $\mu_B$.
A minimum $\chi^2$ fit is performed using data from specific ion-ion (A-A) collisions at a specific beam energy $\sqrt{s_{NN}}$ and centrality.  
The result of the minimum $\chi^2$ fit provides the extracted freeze-out values of $T^{\rm FO},\mu_B^{\rm FO}$ from a given $\sqrt{s_{NN}}$. 
Next, with this pair of $T^{\rm FO},\mu_B^{\rm FO}$ we can calculate the EoS for that specific $\sqrt{s_{NN}}$ and $Y_Q^{\rm{HIC}}$. 
Central collisions are considered the best for such a study since they result in the largest system size and presumably the closest to the infinite volume approximation of the Grand Canonical Ensemble. 

At very large $\sqrt{s_{NN}}$, the nuclei are Lorentz contracted and are very thin along the beam direction, passing through each other nearly instantaneously. As a result, in this limit, there is no time for baryons (i.e.~valence quarks) to be stopped within the collision, such that the global baryon number of the system is extremely small $\langle n_B\rangle \rightarrow 0$. This is also the regime where the collision reaches the highest temperatures. 
However, at lower $\sqrt{s_{NN}}$, the nuclei must be treated as 3D objects, as they are traveling more slowly and take longer to pass through each other, allowing for enough time to capture baryons.  For these intermediate $\sqrt{s_{NN}}$, $n_B$ is large and the initial temperature is lower.  
Finally, at very low $\sqrt{s_{NN}}$, the nuclei may not pass through each other entirely, but rather stick together. In this regime, the initial temperature may be so low that the quark-gluon plasma phase is no longer reached. Thus, the systems created in HICs can vary significantly as a function of $\sqrt{s_{NN}}$ and extend across a wide range on the QCD phase diagram.  

\begin{figure}
    \centering
    \includegraphics[width=\linewidth]{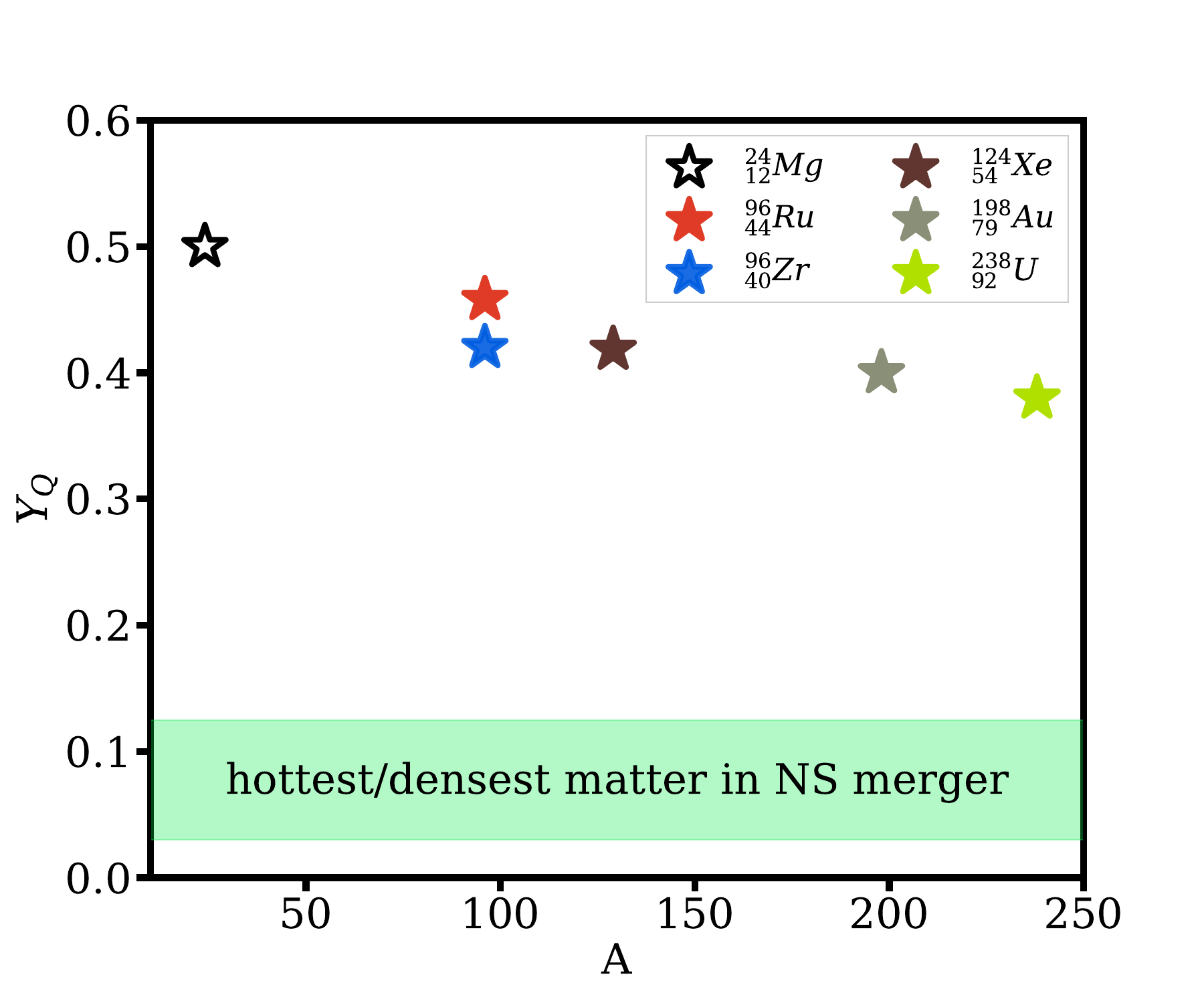}
    \caption{Charge fraction ($Y_Q$) as a function of number of nucleons (A) for different ion species. Filled markers represent ions for which experimental data have already been collected (note that for $^{96}_{44}$Ru and $^{96}_{40}$Zr the data is preliminary). Runs at the LHC with $^{24}_{12}$Mg beams (unfilled, black marker) have been discussed recently but are still in the development stage \cite{lightionworkshop}. The light green band was extracted from Ref.~\cite{Most:2019onn} and indicates the $Y_Q$ range probed by the hottest and densest regions of a neutron star merger throughout the entire merger. There is currently no overlap between the hottest/densest regions in a neutron star merger and the range of $Y_Q$ available in heavy-ion collisions, but more data on different ion species could inform the extrapolation of charge fraction of effects on the EoS. 
    }
    \label{fig:ions}
\end{figure}

Furthermore, different ion species correspond to different $Y_Q^{\rm{HIC}}$.  
In fact, a wide range of $Y_Q^{\rm{HIC}}$ values have already been run at $\sqrt{s_{NN}}=200$ GeV, where the baryon chemical potential is approximately $
\mu_B\sim 20$ MeV (for a summary, see Ref.~\cite{Nana:2024okk}, Table I.). In Fig.~\ref{fig:ions}, we show different ions and their corresponding $Y_Q=Z/A$ and, for comparison, the range of $Y_Q$ reached by the hottest and densest matter in state-of-the-art simulations of neutron star mergers \cite{Most:2019onn}. Note that the range of $Y_Q$ we can currently probe with HICs is limited to $Y_Q \gtrsim 0.4$, much closer to SNM than the neutron-rich matter in neutron star mergers. Thus, we should be cautious when drawing comparisons between theses systems. However, our proposed expansion in Eq.~(\ref{eqn:final_dsdT_YQexpan}) is centered precisely around SNM, and a system scan allows us to extract  $\frac{\partial^2\tilde{S}_{\rm{SNM},2}(T,n_B)}{\partial \delta^2}\big|_{\delta=0}$ from fits of thermal yields. In the future, given enough data across different systems and $\sqrt{s_{NN}}$, it may be possible to extract $\frac{\partial^3\tilde{S}_{\rm{SNM},2}(T,n_B)}{\partial T\partial \delta^2}\big|_{\delta=0}$. Though we highlight, as shown in 
\ref{fig:ions}, that we currently do not have any data for symmetric ions, such as $^{16}$O
or $^{24}$Mg. With the upcoming FAIR fixed target experiment at GSI, there is an opportunity to fill this gap with different $Y_Q^{\rm{HIC}}$,  such that EoS information can be extracted across a range of densities, temperatures, and $\delta$ that would then constrain the coefficients in Eq.\ (\ref{eqn:dsdT_YQexpan}).

Lastly, we expand on a key difference between heavy-ion collisions and astrophysical scenarios (which we first mentioned in Sec.~\ref{subsec:exp_parameter}). In the former, net-strangeness is conserved and, since the initial state has no strangeness, the net-strangeness is guaranteed to be exactly zero. 
Note that does not imply zero strangeness locally, but that the number of strange particles and anti-particles are equal to each other, $N_s=N_{\bar{s}}$ , throughout the collision.
In contrast, in neutron star mergers there is no such conservation, such that the net-strangeness number density may be nonzero.  Thus, if strange degrees of freedom are present at the point of freeze-out, any information about the EoS at a given $(T,\mu_B,\mu_Q)$ extracted from thermal fits of HIC observables corresponds to some curve in $\mu_S$ space, though the net-strangeness density is always zero. In neutron star mergers, $\mu_S$ is exactly zero, but the net-strangeness density may be finite. 
One potential way to avoid this issue is to only study yields, ratios, and fluctuations of light particles such as pions, protons, or light nuclei such as deuterons. 
However, one would have to consider how to map information extracted using the strangeness neutrality constraint in Eq.\ \ref{eqn:str_neu} into the $\mu_S = 0$ plane relevant for neutron star mergers. We also leave that study for a future work. 

\section{Proof-of-principle with a microscopic EOS}\label{sec:pop}

Now that we have constructed and motivated a finite temperature expansion for the dense-matter EoS, we would like to check our proposed framework against a microscopic EoS.   There are a few questions that we wish to explore:
\begin{itemize}

    \item How accurate is the finite $T$ expansion at $\mathcal{O}((T/\mu_B)^2)$ and  $\mathcal{O}((T/\mu_B)^3)$?
    \item Are the terms well-behaved across $Y_Q$? 
    \item Does our expansion in $s/n_B$ accurately describe finite $T$ behavior along different values of $Y_Q$?
    \item Does our switch between a grid in fixed $\vec{\mu}$ to fixed $Y_Q,n_B$ introduce significant numerical error?
\end{itemize}

We study these questions in the context of a relativistic mean field (RMF) model \cite{Alford:2022bpp}, although our approach is generic and can work for any theory not containing first-order phase transitions. This particular RMF model describes protons and neutrons coupled to $\sigma$, $\omega$, and $\rho$ mesons.  
We choose the following values for the nucleon-meson couplings: $g_{\sigma} = 8.3965$, $g_{\omega} = 10.1845$, and $g_{\rho} = 10.9176$ and the meson-meson couplings $b = 0.00438046$, $c =  -0.00359399$, and $b_1 = 5.18964$ (see Ref.~\cite{Alford:2022bpp} for a description of the model Lagrangian and free parameters). These parameters are a representative sample of a large set of constrained parameter values, 
which reproduce the properties of uniform PNM obtained from chiral effective field theory and satisfy basic astrophysical constraints; namely, the maximum mass of a stable, isolated, slowly-rotating neutron star is at least two solar masses \cite{demorest2010two,arzoumanian2018nanograv,cromartie2020relativistic,Fonseca:2021wxt,NANOGrav:2023hde} and the radii predicted for 1.4 and 2.1 solar-mass stars is within the one-sigma posterior obtained from NICER data \cite{Miller:2019cac,Miller:2021qha,Riley:2019yda,Riley:2021pdl}. 

We start by fixing $Y_Q = 0.5$ to study the finite $T$ expansion and demonstrate the accuracy of the finite $T$ expansion at $\mathcal{O}(T^2)$ and $\mathcal{O}(T^3)$. 
Then, we calculate the $s/n_B$ expansion and check its accuracy across $Y_Q$.  
Finally, we quantify the numerical error introduced by switching between a grid in fixed $\vec{\mu}$ to fixed $Y_Q,n_B$, a step that is required for implementing the expanded EoS in simulations of neutron star mergers. 

\subsection{Terms at $\mathcal{O}(T^2)$ and  $\mathcal{O}(T^3)$}

Using the RMF model, we can calculate the $\mathcal{O}(T^2)$ 
term $\partial s/\partial T |_{T=0}$ and the $\mathcal{O}(T^3)$ 
term $\partial^2 s/\partial T^2 |_{T=0}$.  In order to calculate accurate derivatives we require step sizes in the temperature of $\Delta T= 1$ MeV and must generate tables in ranges of $T=0$ to $100$ MeV. 
To do so, we calculate the terms on a fixed grid in $\mu_B$ and $\mu_Q$ for all points at $T=0$.   Then, we can perform a root-finding method to reconstruct $\partial s/\partial T |_{T=0}$ and $\partial^2 s/\partial T^2 |_{T=0}$ along lines of constant $Y_Q$.

\begin{figure}
    \centering
    \includegraphics[width=\linewidth]{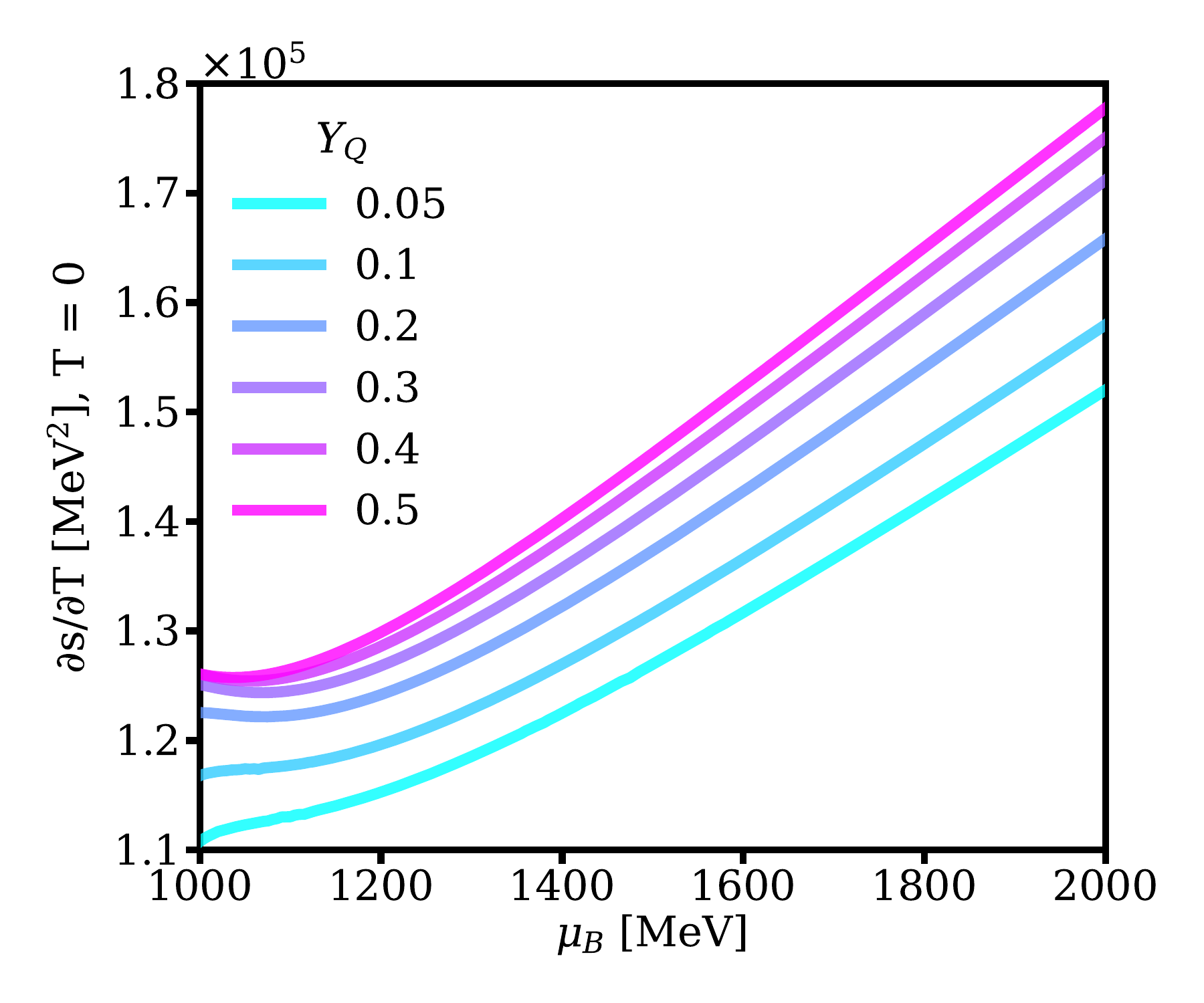}\\
    \includegraphics[width=\linewidth]{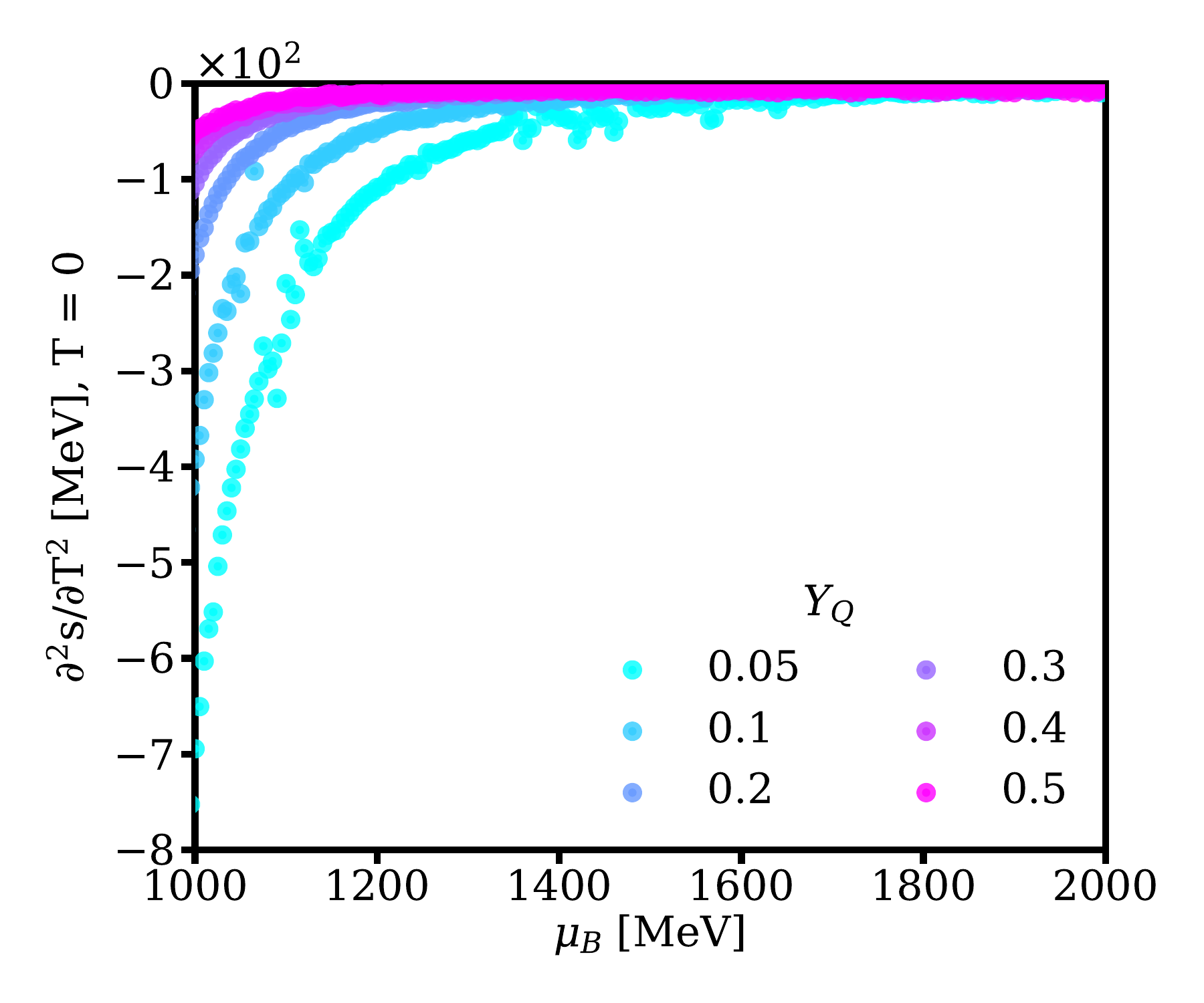}
    \caption{$\partial s / \partial T$ that enters at $\mathcal{O}(T^2)$ (top) and  $\partial^2 s / \partial T^2$ that enters at $\mathcal{O}(T^3)$ (bottom) calculated varying $\mu_B$ for a range of $Y_Q$ values where different colored lines represent different $Y_Q$. }
    \label{fig:dsdT_Yq_dep}
\end{figure}

In Fig.\ \ref{fig:dsdT_Yq_dep} we plot the second-order term $\partial s/\partial T |_{T=0}$ on the top and the third-order term $\partial^2 s/\partial T^2 |_{T=0}$ on the bottom and compare their behavior across $Y_Q$.
Qualitatively, we find a very similar behavior for $\partial s/\partial T |_{T=0}$ across $Y_Q$, but the overall magnitude changes.
When comparing across slices in $Y_Q$, we find that as $Y_Q$ approaches zero,  $\partial s/\partial T |_{T=0}$ decreases.  Thus, from the second-order term alone, we anticipate that SNM is more sensitive to temperature changes compared to PNM. In other words, heavy-ion collisions should be more sensitive to temperature effects than neutron star mergers. The second-order term is positive definite, which is what one should expect, because at finite temperature entropy is non-zero. Thus, $\partial s/\partial T $ at the limit of $T=0$ \emph{must} be positive to ensure that as one goes from $T=0$ to any $T$, entropy is increased.  We find that the $\partial s/\partial T |_{T=0}$ term increases monotonically with $\mu_B$ and has a relatively smooth behavior.  

The third-order term in Fig.\ \ref{fig:dsdT_Yq_dep} (bottom) is negative and approaches zero for increasing $\mu_B$, vanishing more quickly for values of $Y_Q$ closer to SNM.
This term, $\partial^2 s/\partial T^2 |_{T=0}$, is the largest at low $\mu_B$, especially for small $Y_Q$.  
We can conclude that SNM is more sensitive to finite temperature effects, since $\partial s/\partial T $ is largest in that regime, and likely has the most straight-forward temperature dependence because the third-order term is negligible, at least in this RMF model. 
In Sec.\ \ref{sec:error}, we quantify more precisely the influence of the third-order term at specific temperatures. However, in this RMF model,  $\partial^2 s/\partial T^2 |_{T=0}$ is negligible compared to $\partial s/\partial T|_{T=0} $

\subsection{Finite T expansion for symmetric nuclear matter}

\begin{figure}
    \centering
    \includegraphics[width=\linewidth]{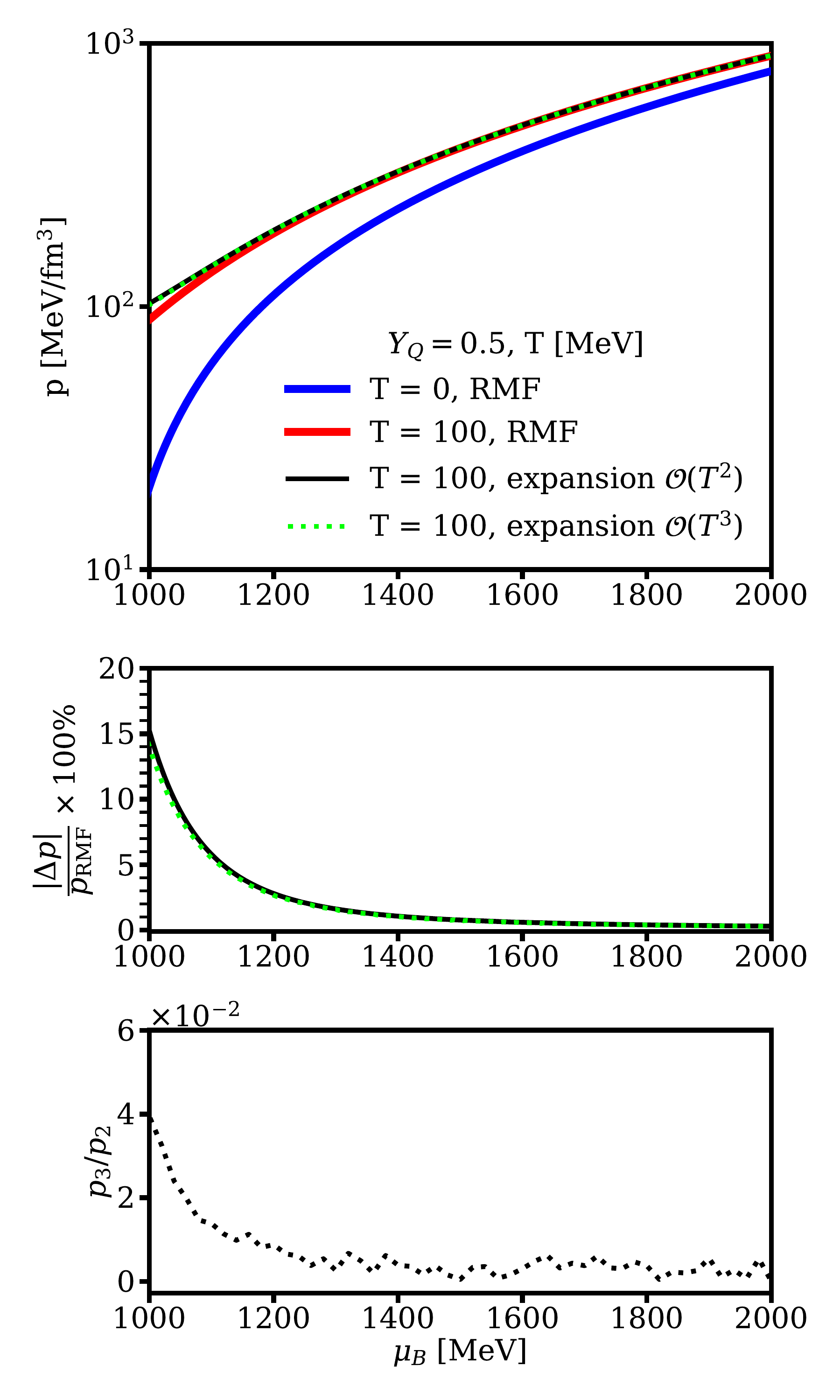}\\
    \caption{The top panel shows the pressure in MeV/fm$^3$ vs baryon chemical potential in MeV at $T=100$ MeV for the original RMF model compared to the reconstructed EOS from the finite $T$ expansion up to different orders. As a comparison, the $T=0$ EOS is shown for the original RMF EOS. 
    The middle panel displays the relative percent error of our expanded pressure vs baryon chemical potential in MeV at $T=100$ MeV compared to the original RMF model. 
    The bottom panel compares the ratio of the third-order over second-order term in Eq.\ (\ref{eq:exp_0}) (that includes the $T$ scaling) vs baryon chemical potential in MeV at $T=100$ MeV. The expansion up to $\mathcal{O}(T^2)$ reproduces the EoS at $T=100$ MeV to with sub-percent error. 
    }
    \label{fig:pT100}
\end{figure}

Now that we know the $\mathcal{O}(T^2)$ and $\mathcal{O}(T^3)$ terms, we can reconstruct the pressure at finite $T$ using Eq.~\ref{eq:exp_1}. 
We use $Y_Q=0.5$, corresponding to SNM, because both  $\partial s/\partial T |_{T=0}(\mu_B)$ and  $\partial^2 s/\partial T^2 |_{T=0}(\mu_B)$ appeared well-behaved in this limit in Fig.\ \ref{fig:dsdT_Yq_dep}.
Further motivation for choosing SNM as a starting point is that we also need to expand around $Y_Q=0$, or $\delta=0$, in order to obtain a $Y_Q$ dependence for  $\partial s/\partial T |_{T=0}(Y_Q)$. 

In Fig.\ \ref{fig:pT100} (top panel) we show a comparison between the pressure $p$ in MeV/fm$^3$ vs $\mu_B$ at $T=0$ and $T=100$ MeV for the RMF EoS. 
Then, our finite $T$ expansion in Eq.\ (\ref{eq:exp_1}) is shown including the second-order term only (black solid line) and both the second- and third-order terms (green dotted line).
We find that the inclusion of the second-order term already provides an extremely accurate representation of the $T=100$ MeV EoS from $\mu_B=1000$ to $2000$ MeV. 
Only a very small deviation can be seen around $\mu_B\sim 1000$ MeV, when $T/\mu_B \sim 0.1$. The negligible difference in the results from including the third-order term was expected from the discussion of Fig.~\ref{fig:dsdT_Yq_dep}.

We can quantify the deviation from the original RMF EoS at $T=100$ MeV by calculating the following relative percent error of deviation from the pressure:
\begin{equation}\label{eqn:relERR}
    \frac{|\Delta p|}{p_{\rm RMF}}\times 100\%=\frac{|p_{\rm expn}-p_{\rm RMF}|}{p_{\rm RMF}}\times 100\% ,
\end{equation}
where $p_{\rm RMF}$ comes directly from the RMF model and $p_{\rm expn}$ comes from our finite $T$ expansion up to order $\mathcal{O}((T/\mu_B)^2)$. The results of this comparison are shown in the middle panel in Fig.\ \ref{fig:pT100}, where we find that for $\mu_B\gtrsim 1200$ MeV our error is only at $\mathcal{O}(1\%)$ and significantly less for higher values of $\mu_B$. 
However, for smaller $\mu_B$ we do find larger deviations from the original RMF table. The larger error in that regime is expected since we see already in Fig.~\ref{fig:dsdT_Yq_dep} that higher order terms become non-negligible at low $\mu_B$ and the expansion parameter is close to 1.

Finally, in the bottom panel of Fig.\ \ref{fig:pT100} we study the contribution of the second-order term,
\begin{equation}
    p_2\equiv\frac{\partial s}{\partial T} \Big|_{T=0} T^2 ,
\end{equation}
against the third-order term,
\begin{equation}
    p_3\equiv\frac{\partial^2 s}{\partial T^2} \Big|_{T=0} T^3 ,
\end{equation}
 by plotting the ratio $p_3/p_2$ vs $\mu_B$.  
 For SNM, the $p_3$ contribution is small and even at its largest it is four orders of magnitude smaller than $p_2$ such that it plays no significant role in the expansion.  

\subsection{$s/n_B$ expansion in $Y_Q$}

Now that we have shown that our finite $T$ expansion works well in the specific limit of SNM, we test if the $s/n_B$ expansion properly captures the $Y_Q$ dependence of $\partial s/\partial T|_{T=0}$.

\begin{figure}
    \centering
    \includegraphics[width=\linewidth]{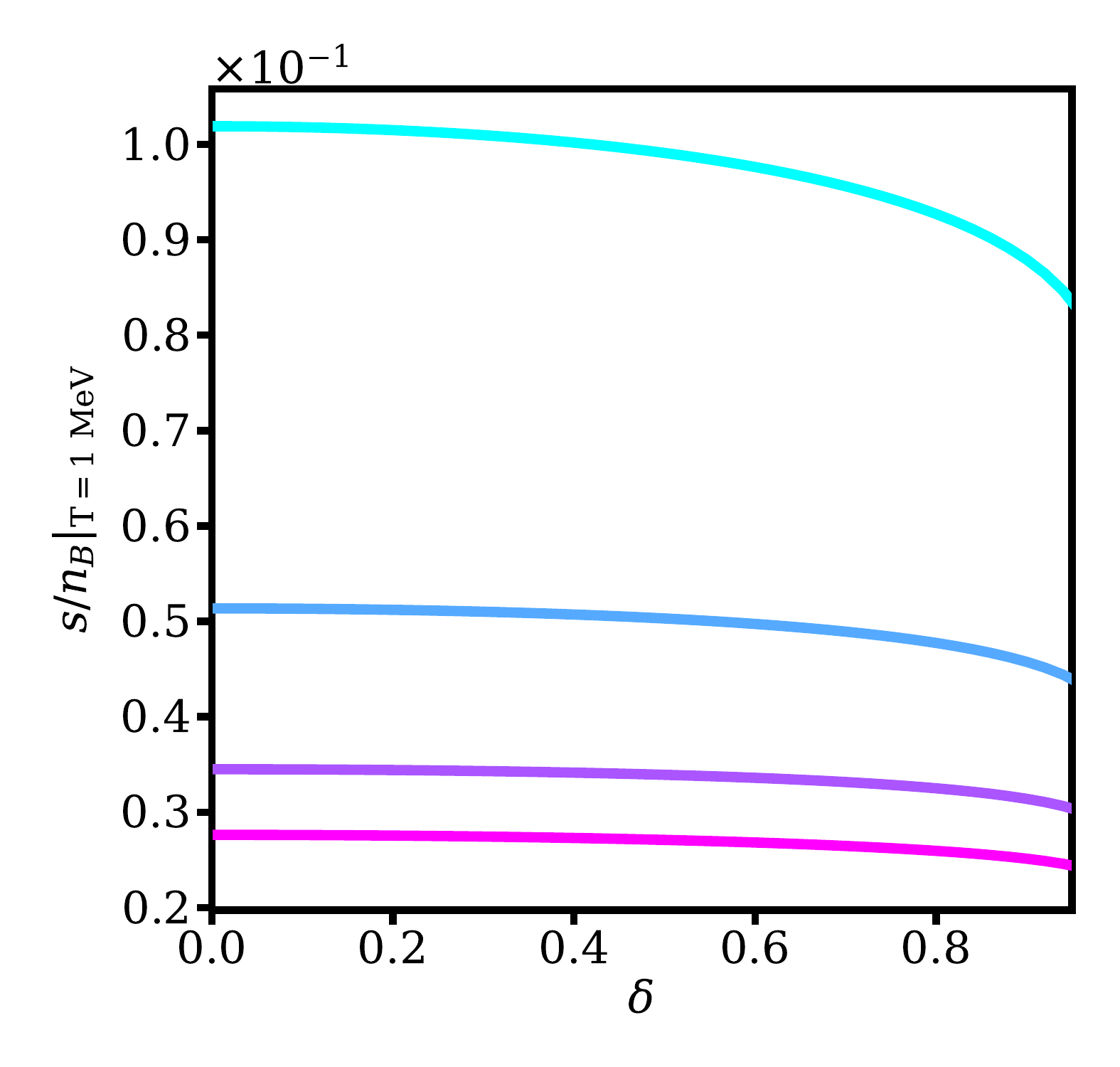} \\
    \includegraphics[width=\linewidth]{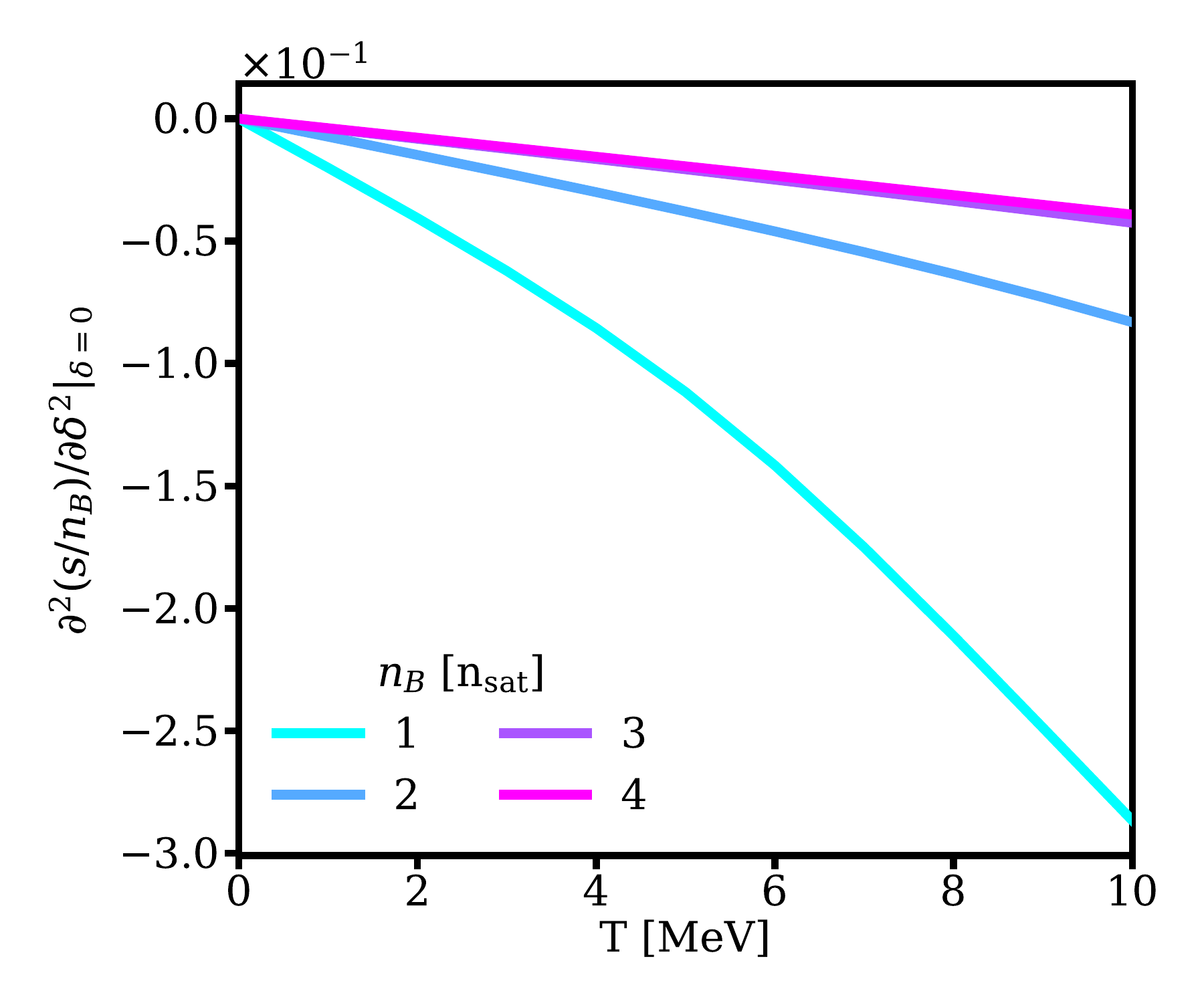}
    \caption{Top: $s/n_B$ versus $\delta$ for different $n_B$ values in terms of the saturation density are shown as lines in different colors. Bottom: The second derivative of $s/n_B$ with respect to $\delta$ is shown at the limit of $\delta=0$ versus the temperature.  Different $n_B$ values in terms of the saturation density are shown as lines in different colors. }
    \label{fig:d2dsnb_ddelta2_across_nB}
\end{figure}

As a first step, we plot $s/n_B$ versus the expansion parameter $\delta$ for a fixed temperature $T=1$ MeV along lines of constant $n_B$.  
The result is shown in the top panel of Fig.\ \ref{fig:d2dsnb_ddelta2_across_nB}. 
We find that $s/n_B$ has a small dependence on $\delta$ for different $n_B$ slices, which becomes more significant for at small $n_B$. 
Though this dependence is small, $s/n_B$ decreases with increasing $\delta$, and it decreases more rapidly for larger values of $\delta$.
Thus, from this behavior we anticipate a negative second derivative in $\delta^2$  that takes on larger values at small $n_B$. 

Next, in Fig.\ \ref{fig:d2dsnb_ddelta2_across_nB} in the bottom panel we show the second derivative, $\partial^2 (s/n_B)/\partial \delta^2|_{\delta=0}$, versus $T$.
We find that $\partial^2 (s/n_B)/\partial \delta^2|_{\delta=0}$ is always negative, which implies that $s/n_B$ decreases with increasing $\delta$, or, in other words, $s/n_B$ is larger (at a fixed $T,n_B$) for SNM than for PNM. 
This qualitative effect was indeed anticipated from the results in the top panel of Fig.\ \ref{fig:d2dsnb_ddelta2_across_nB}.  
The other effect that we see in Fig.\ \ref{fig:d2dsnb_ddelta2_across_nB} (bottom) is that the temperature dependence of $\partial^2 (s/n_B)/\partial \delta^2|_{\delta=0}$ changes significantly with $n_B$ -- for low $n_B$ this term depends strongly on $T$, whereas
at high $n_B$ the temperature dependence is much smaller. This suggests that the $\delta$ dependence of $s/n_B$ is more relevant at lower densities and higher T. 

\begin{figure}[t]
    \centering
    \includegraphics[width=\linewidth]{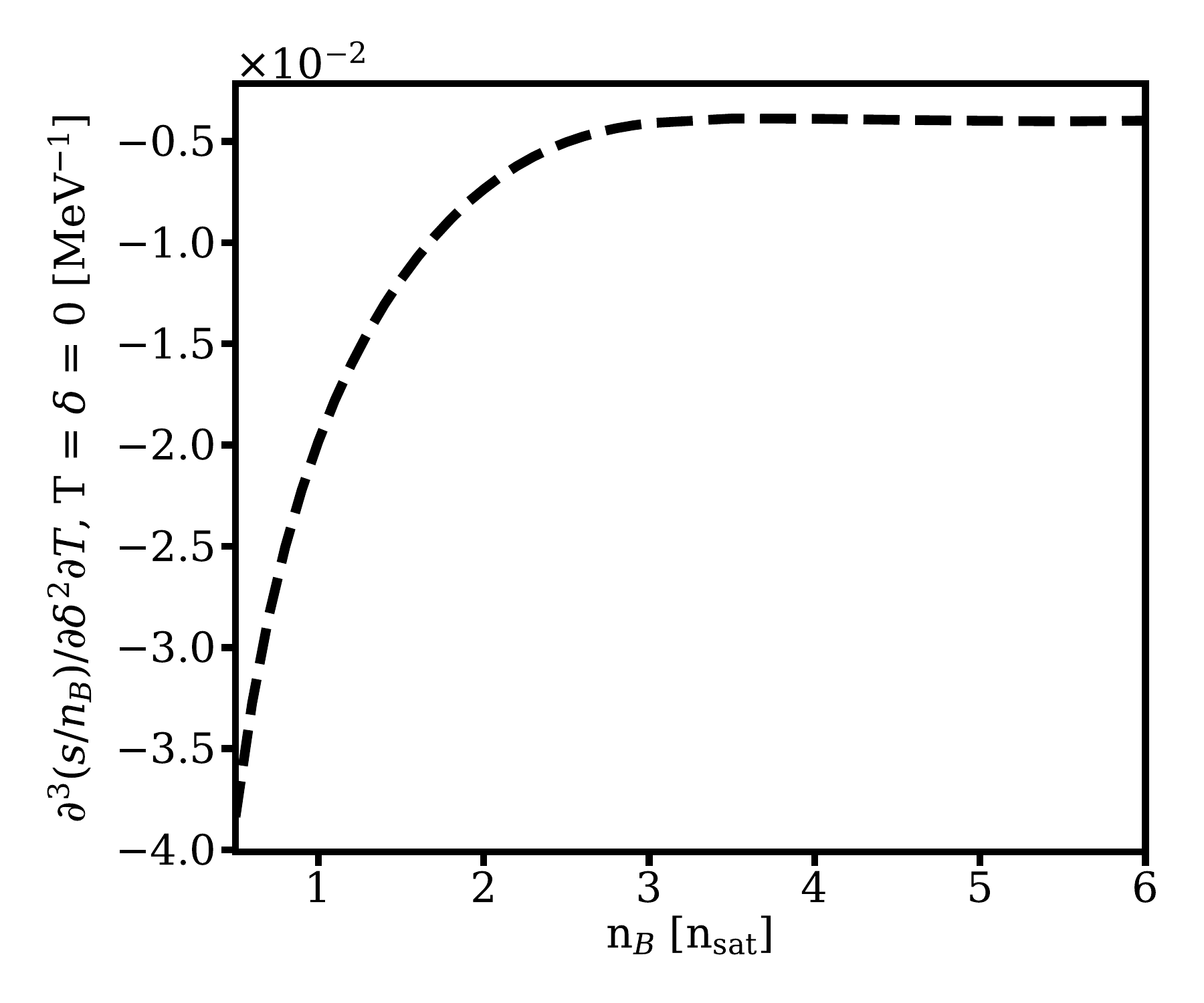} \\
    \caption{The derivative required for the $s/n_B$ expansion (see Sec.\ \ref{sec:yq_effects}) taken in the limit where $T\rightarrow0$ versus baryon density in units of saturation density.}
    \label{fig:d3dsnb_ddelta3_across_nB}
\end{figure}

Finally, we plot in Fig.\ \ref{fig:d3dsnb_ddelta3_across_nB} the third-derivative that is needed for the $s/n_B$ expansion in Eq.\ (\ref{eqn:dsdT_YQexpan}).  
The overall term is negative as we would expect from the combination of Fig.\ \ref{fig:dsdT_Yq_dep} and Eq.\ (\ref{eqn:dsdT_YQexpan}).
Fig.\ \ref{fig:dsdT_Yq_dep} demonstrates that $\partial s/\partial T|_{T=0}$ must decrease with decreasing $Y_Q$, which is only possible from Eq.\ (\ref{eqn:dsdT_YQexpan}) if the term $\partial^3 (s/n_B)/\partial \delta^2\partial T|_{\delta=T=0}$ is negative. 
Additionally, we find that the overall magnitude of $\partial^3 (s/n_B)/\partial \delta^2\partial T|_{\delta=T=0}$ is the largest for small $n_B$, such that small $n_B$ is be most sensitive to changes in $Y_Q$.
The fact that low $n_B$ is more sensitive to $Y_Q$ changes is another source of potential error that may appear at low $n_B$.  

\begin{figure}
    \centering
    \includegraphics[width=\linewidth]{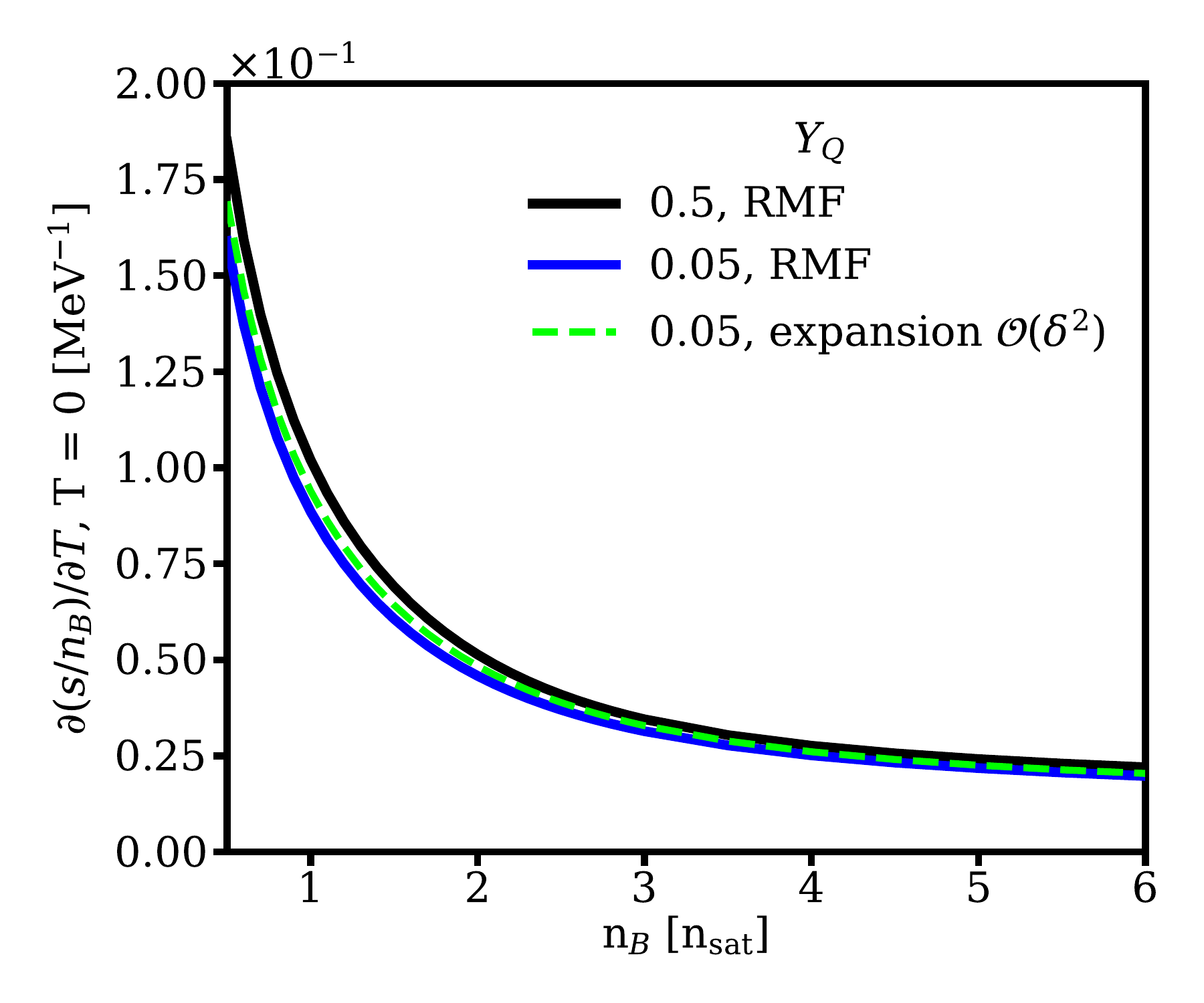}
    \caption{$\partial (s/n_B) / \partial T |_{T=0}$ as a function of $n_B \;[n_{\rm sat}]$. The black solid line is the reference $\delta = 0$, or $Y_Q = 0.5$, about which we expand. The blue solid line is the true value from the RMF model at $Y_Q=0.05$. The green dashed line is the approximation from $s/n_B$ expansion up to $\mathcal{O}(\delta^2)$.}
\label{fig:isentrope_expansion_test_Yq001}
\end{figure}

We can test how well our expansion works by studying how $\partial (s/n_B) / \partial T |_{T=0}$ varies with $Y_Q$ in the RMF model and then try to reproduce those results for a small value of $Y_Q$ using Eq.\ (\ref{eqn:dsdT_YQexpan}).  
The result of the $s/n_B$ expansion at finite temperatures is shown in Fig.\ \ref{fig:isentrope_expansion_test_Yq001} where we used SNM as our starting point and expanded to $Y_Q=0.05$ for a baryon density range of $n_B=0.5 - 6\; n_{\rm sat}$. 
The difference between $\partial (s/n_B) / \partial T |_{T=0}$ for $Y_Q=0.5$ and $Y_Q=0.05$ is already very small and only shows small deviations at low $n_B$, whereas for high $n_B$ there is almost no $Y_Q$ dependence, as anticipated from our discussions of Figs.~\ref{fig:d2dsnb_ddelta2_across_nB}- \ref{fig:d3dsnb_ddelta3_across_nB}. 
The prediction from the $s/n_B$ expansion for $Y_Q=0.05$ is compared to the value evaluated directly from the RMF model, and the  $s/n_B$ expanded results reproduce the behavior of $\partial (s/n_B) / \partial T |_{T=0}$ vs $n_B$ very well. 
At low $n_B$, we do see a small deviation, but this regime also corresponds to low $\mu_B$, such that we already anticipate more numerical error there from the finite temperature expansion itself. 

\subsection{Finite temperature expansion for a range of $Y_Q$}

In the previous two subsections, we established the accuracy of a finite temperature expansion Eq.~(\ref{eq:exp_1}) for SNM and an expansion of $(\partial s/\partial T)_{T=0}$ centered around SNM Eq.~(\ref{eqn:dsdT_YQexpan}). We now check the error that is introduced by combining these two procedures, such that a 3D EoS can be obtained using as a starting point the EoS of SNM at $T=0$ (Eq.~(\ref{eqn:final_dsdT_YQexpan})).
\begin{figure}[]
    \includegraphics[width=.9\linewidth]{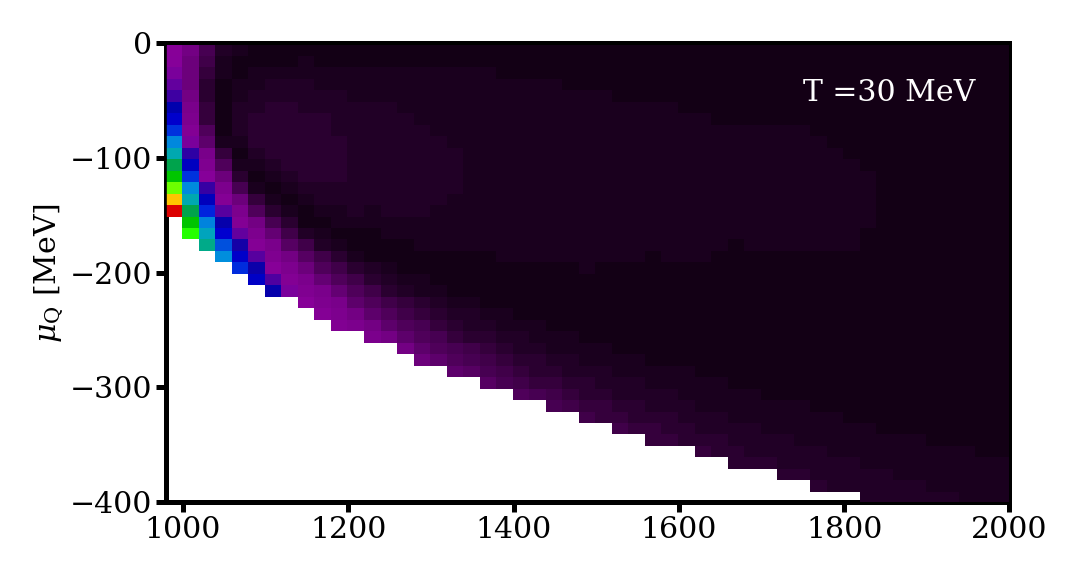} \\
    \hspace{3mm}\includegraphics[width=0.9\linewidth]{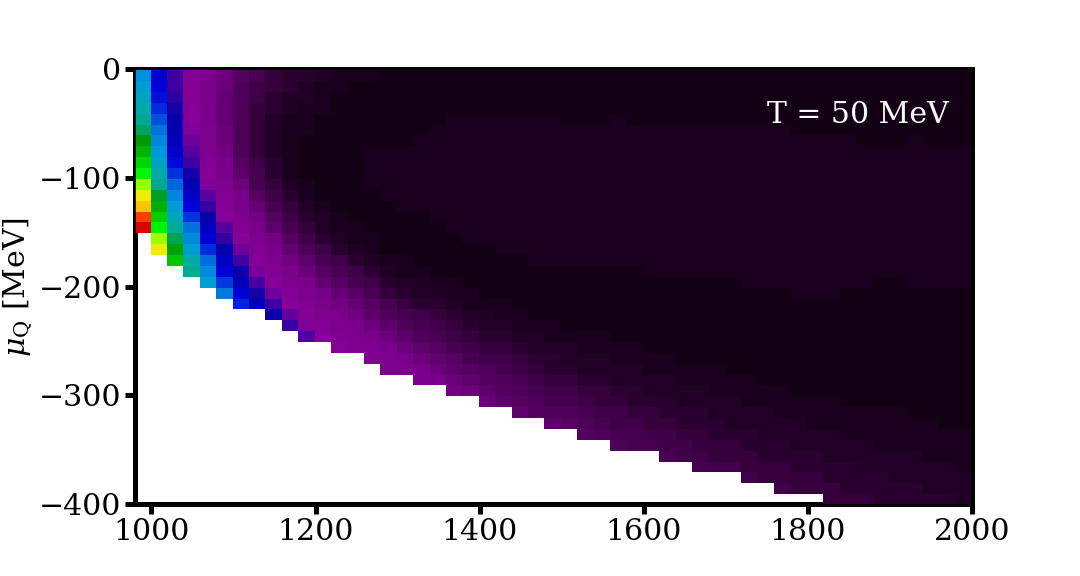}\\
    \centering\includegraphics[width=.9\linewidth]{ 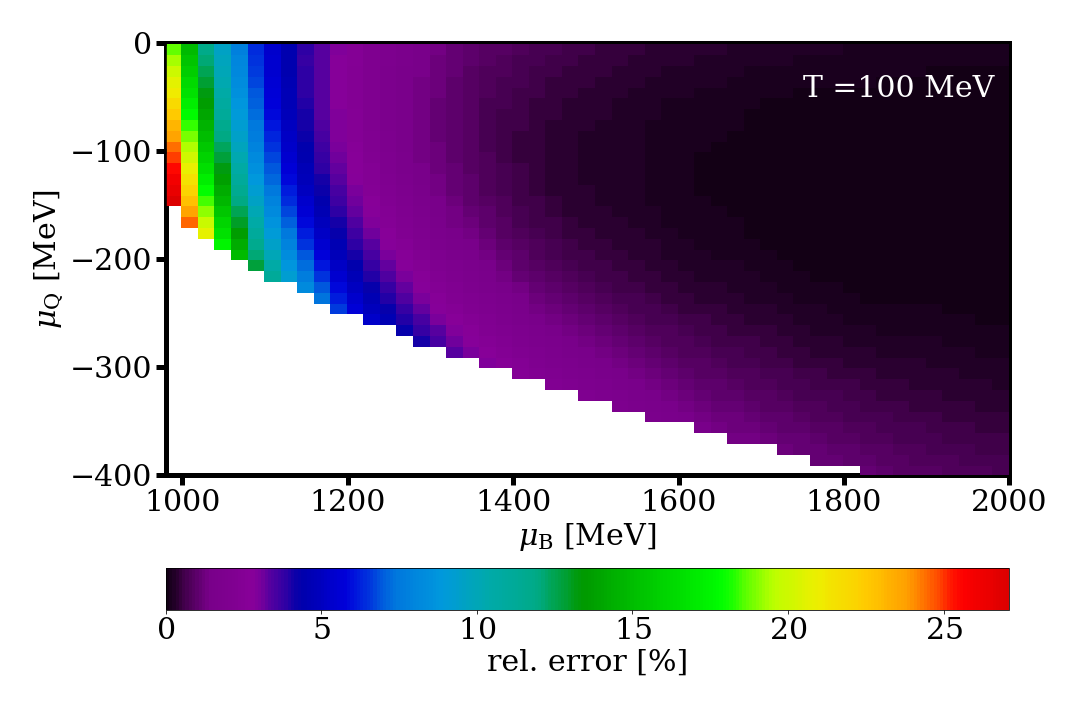}
    \caption{
    The relative error percentage of the pressure, see Eq.\ (\ref{eqn:relERR}), is shown across the range of $\mu_B,\mu_Q$ relevant to neutron star mergers. The top panel has a temperature of $T=30$ MeV, the middle panel has a temperature of $T=50 $ MeV, and the bottom panel has a temperature of $T=100$ MeV.}
    \label{fig:error_estimation}
\end{figure}

We summarize our findings in Fig.\ \ref{fig:error_estimation}. We performed an error quantification for the expansion from Eq.\ (\ref{eqn:final_dsdT_YQexpan}) across the $\mu_B,\mu_Q$ plane at three different temperatures $T=30,50,100$ MeV, shown in the top, middle, and bottom panels, respectively.
We show density plots across $\mu_B,\mu_Q$ where the colors indicate the $\frac{|\Delta p|}{p_{\rm RMF}}\times 100\%$ as defined in Eq. (\ref{eqn:relERR}).
Our results for $T=30$ MeV suggest that most of the error is less than $1\%$, although some higher deviations appear around $\mu_B\sim 1000$ MeV.  We also do not see significant $\mu_Q$ dependence of the error. 
A similar picture is seen at $T=50$ MeV although larger deviations begin to appear around $\mu_B\sim 1100$ MeV but higher $\mu_B$'s also display error only at the $\sim 1\%$ level.
Even up to $T\sim 100$ MeV our expansion still works quite well above  $\mu_B\gtrsim 1150$ MeV, where we only see deviations at the percent level. 
However, it is clear that between $1000 \lesssim \mu_B\lesssim 1150$ MeV and $T= 100$ MeV that the error can be as high as $15-20\%$, consistent with our findings in Fig.\ \ref{fig:pT100}.

\subsubsection{Conversion between a fixed grid in  $\vec{\mu}$ vs a fixed grid in $n_B,Y_Q$}

\begin{figure}[h]
    \centering
    \includegraphics[width=\linewidth]{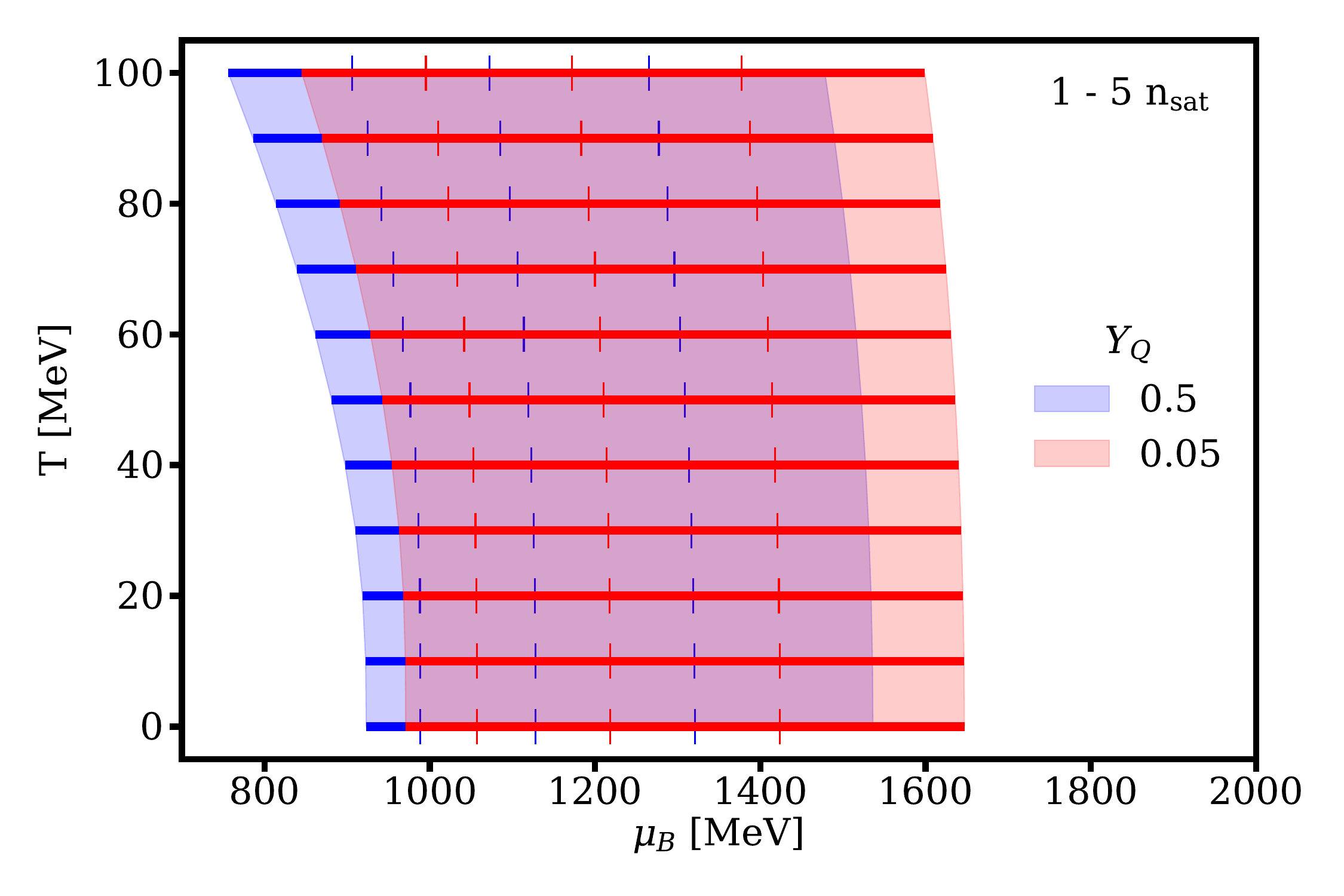}
    \caption{Comparison of the range in $\mu_B$ covered for the baryon density range of $1-5 \ n_{\rm sat}$ at different slices in temperature $T$. Integer units of $n_B [n_{\rm sat}]$ are indicated by the minimum and maximum of each region and tick marks. Both $Y_Q=0.5$ and $Y_Q=0.05$ are shown. The results come directly from the RMF model.}
    \label{fig:nB_muB}
\end{figure}

It is significantly more natural thermodynamically to perform our finite $T$ expansion across a grid of fixed $(T,\vec{\mu}$)   than across a grid of $(T,n_B,Y_Q)$, even though the latter are more natural variables in hydrodynamical simulations.  
We now discuss the consequences of performing our expansion in $(T,\vec{\mu})$ and then the need to regrid the EoS in terms of $(T,n_B,Y_Q)$ so that it can be used in numerical relativity simulations. 

We saw already in Fig.\ \ref{fig:pT100} that in the range of $1000 \lesssim \mu_B \lesssim 2000$ MeV the finite $T$ expansion can reproduce the pressure extremely well at $T=100$ MeV, although small deviations begin to appear at $\mu_B\lesssim 1150$ MeV. 
One complication to this picture, however, is that a given range of $n_B$ has a non-trivial relationship with $\mu_B$ at finite $T$.  
In other words, if we take the range of $\mu_B=1000-2000$ MeV and then calculate the baryon density at different temperatures, we find that
\begin{equation}
    n_B(T=100 ,\mu_B=1000)\neq n_B(T=0 ,\mu_B=1000) ,
\end{equation}
where the units are in MeV.
While this may seem an obvious statement, it has important consequences for error quantification in our approach. 

Since the error quantification in Fig.\ \ref{fig:pT100} was done at fixed $\mu_B$, the corresponding $n_B$ for that range of $\mu_B$ is not the same at fixed $T$.  
Rather, as we go to higher $T$, we actually require a \emph{lower} range in $\mu_B$ in order to produce the same range in $n_B$. 
To see this effect clearly, we fix our range in $n_B=1-5\ n_{\rm sat}$, which is a reasonable estimation of the range in $n_B$ for most of the matter produced in neutron star mergers (see, e.g., \cite{Most:2018eaw,Most:2019onn,Perego:2019adq}). 

In Fig.\ \ref{fig:nB_muB}, we plot the interval in $\mu_B$ corresponding to $n_B=1-5\ n_{\rm sat}$ at different temperatures for $Y_Q=0.5$ (blue shaded region) and $Y_Q=0.05$ (red shaded region), and indicate integer units of $n_B [n_{\rm sat}]$ by the minimum of each shaded region (corresponding to 1 $n_{\rm sat}$), tick marks, and the maximum of each shaded region (5 $n_{\rm sat}$).  We find that the fixed range in $n_B=1-5n_{\rm sat}$ shifts to \emph{lower} $\mu_B$ for higher $T$.
The implication of this effect is that for a fixed $n_B$ the error increases with increasing $T$ not only from the expansion itself but also because of the shift to lower $\mu_B$ where the error is larger. 
Furthermore, this shift is stronger for SNM (i.e., $Y_Q=0.5$) than for neutron-rich matter in neutron stars (shown for $Y_Q=0.05$).  
We note, however, that this effect is not that large at temperatures of $T\lesssim 50$ MeV, so depending on the maximum $T$ reached within a merger it may play a less significant role. 

We can also see this effect numerically using the expansion derived in Appendix \ref{app:thermoseries} in Eqs.\ (\ref{eqn:nBoT3},\ref{eqn:nBexpan}). 
If we just take the expansion of $n_B$ up to $\mathcal{O}(T^2)$ shown in Eq.\ (\ref{eqn:nBexpan}), we find that the change of $n_B$ with $T$ at a fixed $\vec{\mu}$ can be determined as:
\begin{equation}
    n_B(T,\vec{\mu})- n_B ( \vec{\mu}) \biggr\rvert_{T=0}=\frac{1}{2}\frac{\partial^2 s(T,\vec{\mu})}{\partial T \partial \mu_B}\biggr\rvert_{T=0, \mu_Q} T^2 ,
\end{equation}
where the sign of $\frac{\partial^2 s(T,\vec{\mu})}{\partial T \partial \mu_B}\bigr\rvert_{T=0, \mu_Q}$ tells you if $n_B$ increases or decreases with $T$.
However, in Fig.\ \ref{fig:dsdT_Yq_dep} (top) we can already get a sense of sign of this derivative because we see that $\frac{\partial s(T,\vec{\mu})}{\partial T }\bigr\rvert_{T=0,Y_Q}$ monotonically increases with $\mu_B$ such that the derivative is most likely always positive (note that $Y_Q$ is constant, rather than constant $\mu_Q$ needed for our expansion, however, we expect that difference to be small). 
This is consistent with Fig.~\ref{fig:nB_muB}, where we found that at a fixed $\vec{\mu}$, $n_B$ should always increase with T.

\begin{figure}[t]
    \centering
    \includegraphics[width=\linewidth]{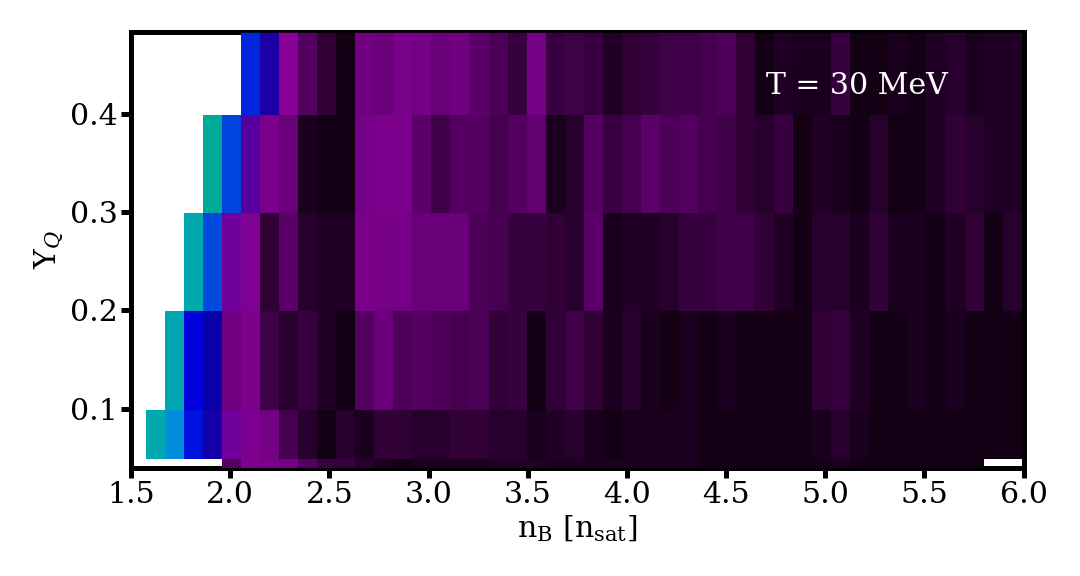} \\
    \includegraphics[width=\linewidth]{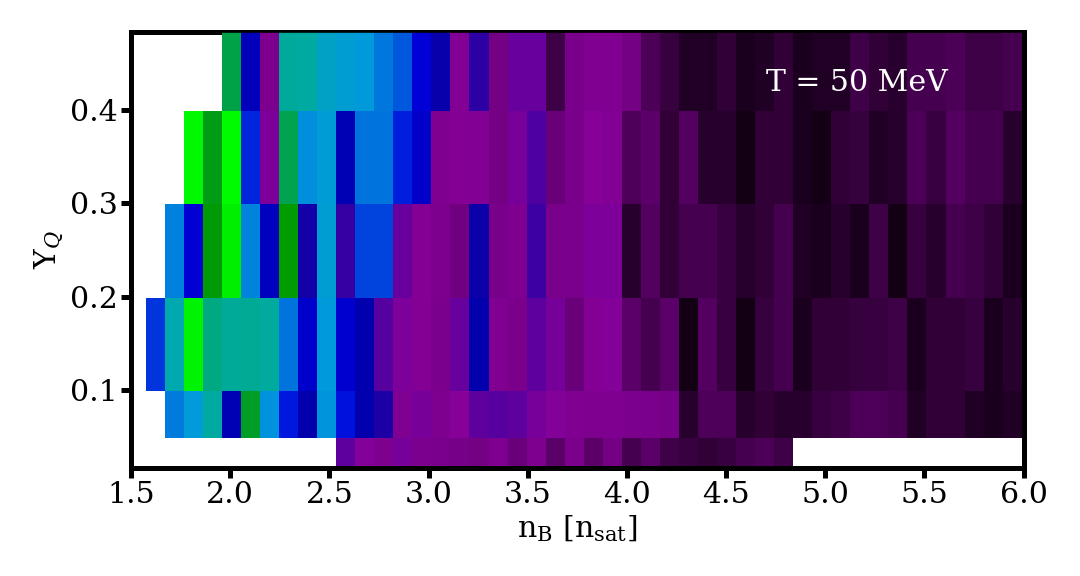}\\
   \includegraphics[width=\linewidth]{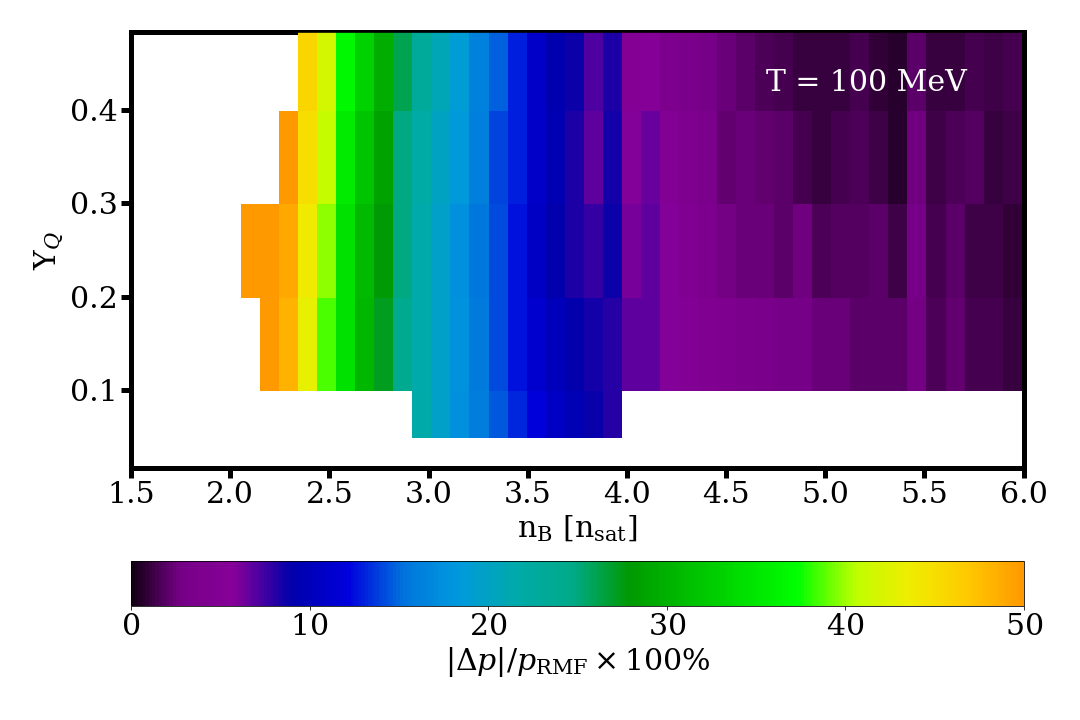} \\
    \caption{The relative error percentage of the pressure, see Eq.\ (\ref{eqn:relERR}), is shown across the range of $n_B,Y_Q$ relevant to neutron star mergers. The top panel has a temperature of $T=30$ MeV, the middle panel has a temperature of $T=50$ MeV, and the bottom panel has a temperature of $T=100$ MeV.
    }
    \label{fig:yq_nb_error}
\end{figure}

With this in mind, we can then understand the relative error percentage of the pressure as a function of $n_B,Y_Q$ shown in Fig.\ \ref{fig:yq_nb_error}. 
To make Fig.\ \ref{fig:yq_nb_error}, we calculate thermodynamic derivatives, such that numerical error at low $\mu_B$ can influence regions of higher $\mu_B$. 
Thus, we find that in Fig.\ \ref{fig:yq_nb_error} there is a clear increase in the relative error percentage of the pressure across $n_B,Y_Q$.
Above $n_B\gtrsim 3 n_{\rm sat}$ the error still remains small and is under $10\%$ even up to $T=100$ MeV.
Similar to the behavior across $\mu_Q$ in Fig.\ \ref{fig:error_estimation}, we do not see almost any $Y_Q$ dependence to the error in  Fig.\ \ref{fig:yq_nb_error}.

In order to better understand the low $n_B$ error, we compare the two thermodynamic bases at $T=30$ MeV. In Fig.\ \ref{fig:error_estimation} for $\mu_B\gtrsim 1000$ MeV most of the error is at the percent level. 
Using our results from Fig.~\ref{fig:nB_muB}, we know that $T=30$ MeV and $\mu_B\gtrsim 1000$ corresponds to $n_B\sim 1.5 n_{\rm sat}$ (there is little shift in the $n_B(\mu_B)$ relationship at that low of a $T$). 
However, the numerical error in Fig.\ \ref{fig:yq_nb_error} for $T=30$ MeV and $n_B\sim 1.5 n_{\rm sat}$ is quite large varying from $10-20\%$. 
Thus, we can conclude that this increase arises primarily from the numerical derivatives rather than the mapping between thermodynamic bases. 

\section{Error Quantification and Future Improvements}\label{sec:error}

\begin{figure}[t]
    \centering
    \includegraphics[width=\linewidth]{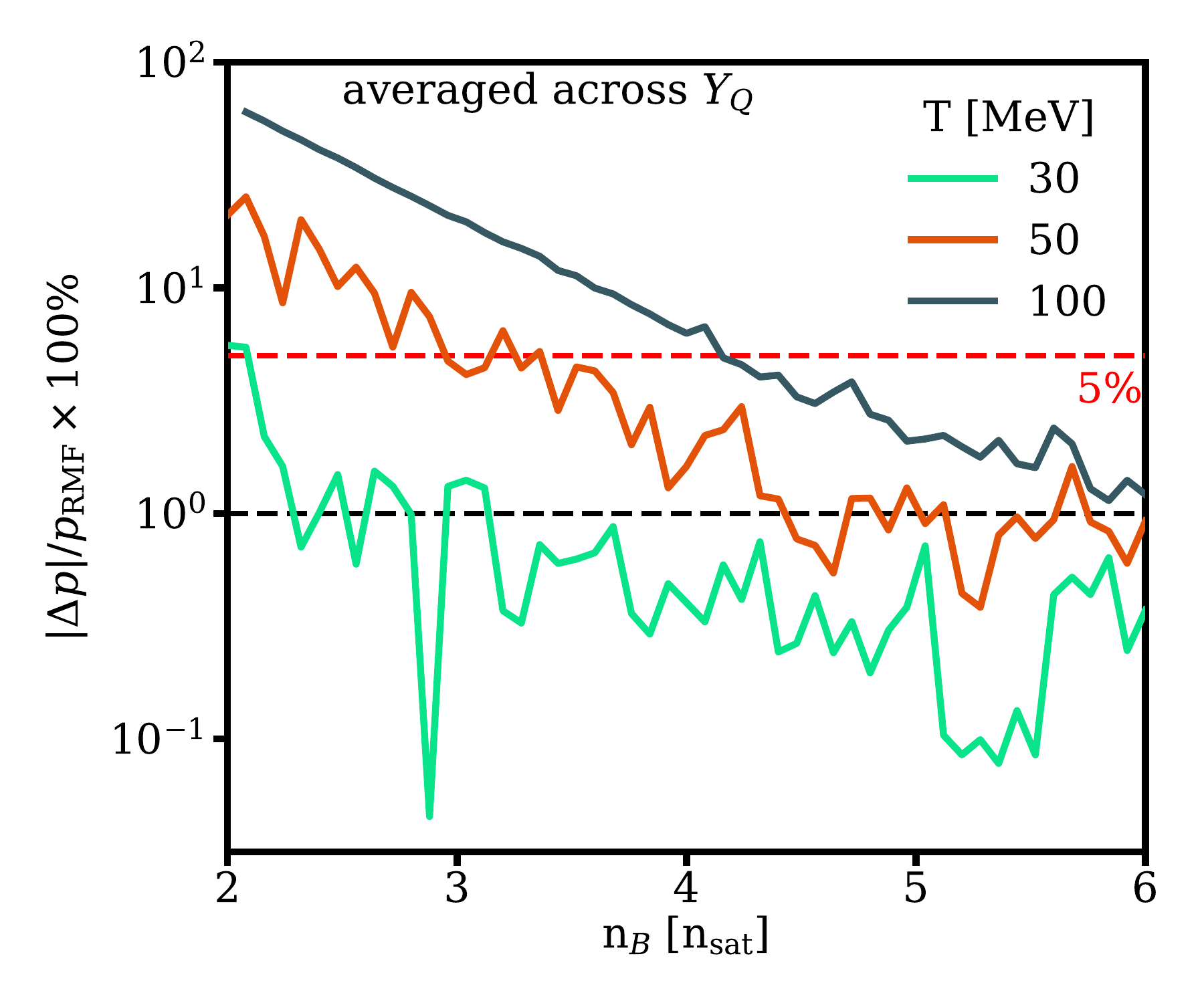} \\
    \caption{Relative percentage error in the pressure, see Eq.\ (\ref{eqn:relERR}), from the finite temperature and $s/n_B$ expansions, see Eq.~(\ref{eqn:final_dsdT_YQexpan}), as a function of $n_B$ in $n_{\rm sat}$, averaged across $Y_Q\in [0.02,0.2]$ (most relevant range for neutron star mergers) for $T=$ 30, 50, and 100 MeV. The compounded error in $n_B$ and $Y_Q$ decreases the accuracy of the expansion on the $n_B,Y_Q$ plane compared to $\mu_B,\mu_Q$ at a fixed $T$.
    }
    \label{fig:yq_averaged_error}
\end{figure}

In this section, we break down the different sources of error in our approach and methods that can be used in future work to improve the error. 
To be clear, we will begin with the assumption that the SNM EoS is known and only discuss error from the expansion itself. 
Additionally, we should clarify that there are three different types of error that arise in our approach:
\begin{itemize}
    \item {\bf Numerical Error.} The first source of error is the most intuitive to understand. We require a number of derivatives in our approach, some that are even up to third-order. An especially significant challenge is that these derivatives are often temperature derivatives that must be taken in the limit of $T=0$. However, a known issue in microscopic EOS codes is that the Fermi-Dirac distribution is difficult to solve in this regime, leading to numerical noise. Another potential source of numerical error is taking derivatives close to the edge of a grid.
    \item {\bf Expansion Error.} We use two different expansions in our approach and only consider terms up to $\mathcal{O}(T^2)$ for the finite $T$ expansion, $\mathcal{O}(\delta^2)$ for the $s/n_B$ expansion. Error can arise from higher-order contributions in these expansions.
    \item {\bf Uncertainty in coefficients.} The input for our 3D expansion requires knowledge of two, density-dependent functions. These can be obtained from experiments or be calculated from a given microscopic EoS, both would introduce additional uncertainty in the calculation.
\end{itemize}

\subsection{Numerical Error}\label{sec:numerical_err}

The numerical error in our approach is a relatively small contribution.
The primary issue is that it is difficult to calculate the entropy accurately in the RMF model at low $T$, which is required for our second and third-order terms in Fig.\ \ref{fig:dsdT_Yq_dep}.  
There are ways to improve these calculations in microscopic models and we plan to explore them in a future work. 
However, considering how well our $\mathcal{O}(T^2)$ results reproduce the pressure up to $T=100$ MeV in Fig.\ \ref{fig:pT100} and Fig.\ \ref{fig:error_estimation}, we believe that the numerical error at least in $\partial s/\partial T|_{T=0}$ is likely quite small on a grid of $\vec{\mu}$. 

Another source of numerical error arises after the 3D pressure is calculated and we obtain the other thermodynamic observables in the EoS such as $s$, $\varepsilon$, $n_B$, $Y_Q$.  
We must take first-order numerical derivatives of the pressure to recover these thermodynamic variables or calculate these analytically with further information about $T=0$ (see Appendix \ref{app:thermoseries}). 
In this work, we calculated the thermodynamic variables numerically and find a significant increase in the error when switching to a grid of $n_B,Y_Q$. 
The numerical error arises for a few different reasons: the shift in $n_B$ to lower $\mu_B$ at finite $T$, the higher error from the expansion up to $\mathcal{O}(T^2)$ in that regime, and potentially boundary effects when taking numerical derivatives since the error is most pronounced close to the edge of our grid.

\subsection{Expansion Error}

We first discuss error that can arise in keeping terms in the finite $T$ expansion only up to $\mathcal{O}(T^2)$.  

We define the order-of-magnitude of the \emph{maximum} contribution to the second and third-order terms based on the results shown in Fig.~{\ref{fig:dsdT_Yq_dep}},
\begin{eqnarray}
    \frac{\partial s}{\partial T}\bigg|_{T=0}&\sim& 10^5 ,\nonumber \\
     \frac{\partial^2 s}{\partial T^2}\bigg|_{T=0}&\sim& 10^2.\nonumber
\end{eqnarray}
We should note that these are dimensionful terms such that they should be considered in conjunction with the relevant power of $T$.  Then, we can take the ratios of the second and third-order terms including their $T$ contributions, i.e.,
\begin{eqnarray}
    \frac{\frac{\partial s}{\partial T}\big|_{T=0} T^2}{\frac{\partial^2 s}{\partial T^2}\big|_{T=0} T^3}&\sim& \frac{10^2\; T^3}{10^5\; T^2}\nonumber\\
    &=& 10^{-3}\;T.
\end{eqnarray}
Thus, for a contribution from the third-order term on the order of $1\%$, one requires a temperature of at least $T\sim 10$ MeV.  For a contribution on the order of $10\%$, one requires a temperature of at least $T\sim 100 $ MeV. 
Although, we remind the reader here that these are the \emph{maximum} contributions from the third-order terms and for other regions of the EoS the contribution is sometimes orders of magnitude smaller. 
Typical neutron star mergers reach up to temperature of about $T\sim 50$ MeV but some simulations find temperatures up to $T\sim 100$ MeV \cite{Most:2018eaw,Most:2019onn}. 

We also provide a word of caution here. In the RMF model, we consider protons and neutrons only and no phase transitions exist between these hadrons and quarks. 
The order-of-magnitude or the behavior of these contributions may change for models that have more complex phase structures. 

We also have the $s/n_B$ expansion across $\delta$. 
We found in this work that works very well with terms up $\delta^2$, though the low $n_B$ regime appears to be most sensitive to changes in $\delta$.  It may be possible to explore higher terms up to $\delta^4$ in future work but this appears to be one of the smallest sources of error in this work, since the $\delta$ dependence of finite temperature effects is already small. 

\subsection{Uncertainty in coefficients}

We have the following input functions of $n_B$,  
\begin{itemize}
    \item  $\dfrac{\partial s(T,n_B,Y_Q=0.5)}{\partial T}\Bigg|_{T=0} (n_B)$,
    \item $\dfrac{\partial^3\tilde{S}_{{\rm sym},2}(T,n_B)}{\partial T\partial \delta^2}\Bigg|_{T=\delta=0} (n_B)$,
\end{itemize}
that were first introduced in this work and, therefore, have not yet been constrained from theoretical and experimental work.  
However, since it is possible to extract information about $s/n_B$ at freeze-out from heavy-ion collisions, and we have the flexibility to vary $Y_Q^{\rm HIC}$ by changing $Z/A$, there is hope for experimental information on these coefficients in the future.
Furthermore, we argue that work is warranted on better understanding these functions using a variety of microscopic models. 
For the moment, we have assumed that these functions are known from a model and quantified the error under that assumption.  

\subsection{Final error quantification}

In Fig.\ \ref{fig:yq_averaged_error} we summarize our findings for the error quantification across $n_B$ at different slices in the temperature $T$.  
The error is averaged over the relevant range of $Y_Q$ for neutron star mergers, i.e. $Y_Q=0.02 - 0.2$, and shown on a log scale to emphasize the accuracy across $n_B$. 
We find an overall trend of the error decreasing with $n_B$, consistent with previous plots. 
Using $5\%$ and $10\%$ error as a gauge, we find that the lower temperatures of $T=30$ MeV reach $5\%$ error at $n_B\sim 2 \ n_{\rm sat}$, whereas temperatures of $T=50$ MeV reach $5\%$ error at $n_B\sim 3 \ n_{\rm sat}$. For even higher temperatures of $T=100$ MeV, the error is above $5\%$, but below $20\%$ for densities above 2 $n_{\rm sat}$. Note that the error that appears in Fig.\ \ref{fig:pT100} is not the same as what appears in Fig.\ \ref{fig:yq_averaged_error}, since in Fig.\ \ref{fig:pT100} we show the error on a grid of $\mu_B$ for SNM, and Fig.\ \ref{fig:yq_averaged_error}
 shows the average error for much lower values of $Y_Q$, after both expansions have been applied and the grid has been converted to $(T,n_{\rm sat},Y_Q)$.
 
In analyzing the error shown in Fig.~\ref{fig:yq_averaged_error} and prospects for future improvement,  we consider the following points:
\begin{enumerate}
    \item The finite $T$ expansion up to $\mathcal{O}(T^2)$, shows the largest deviations at low $\mu_B$. This error can be improved once entropy at low $T$ is numerically more accurate within the RMF model, allowing us to check contributions from higher-order terms in the series. 
    Recent work has been proposed \cite{Gholami:2025cfq} to improve numerical accuracy of derivatives in such models.
    \item The $s/n_B$ expansion up to $\mathcal{O}(\delta^2)$ shows the largest deviations from model calculations at low $n_B$. Since the error is nearly independent of $Y_Q$ at a fixed $n_B$, this error is likely negligible. 
    \item A fixed value of $n_B$ shifts to lower $\mu_B$ at finite $T$. Because the expansion is worse at lower $\mu_B$ then this shift means that at finite $T$ there is a larger $n_B$ regime with high error.  This is technically not an error that can be improved on, but rather a physical consequence of $n_B$ at finite $T$.  However, improvements in 1. will help to reduce the error that arise from this issue.
    \item Calculation of numerical derivatives, e.g.~$s,\ n_B,\ n_Q$ at finite $T$. There are multiple challenges taking these derivatives as discussed Sec.\ \ref{sec:numerical_err}. Future work can explore an analytical method to mitigate this issue, using the expansions derived in Appendix \ref{app:thermoseries}.
\end{enumerate}

Of these four sources of error, working to resolve 1. would most likely have the largest impact on the uncertainty quantification because it would also aid with 3. and might make it possible to calculate the derivatives in 4. analytically. 
Additionally, we argue that improving upon 1. should be prioritized because our largest uncertainties occur at low $n_B$ and they are in part driven by 1. 
The best starting point to fix this error is to improve the entropy calculations in the RMF model at low $T$ and studying the influence of higher-order contributions in $T$.

We also argued that 4. is a significant source of error in this work.  In the future, we could improve on this issue by creating a grid in $\vec{\mu}$ well-beyond the regime of validity of the model to avoid boundary issues and trying other numerical approaches to calculate these derivatives. 
There is also the possibility of analytically calculating these state variables, but, as previously stated, that requires knowledge of new, potentially challenging derivatives at $T=0$ that have not yet been calculated. 

In contrast, the error arising from expansion in $\delta$ is the smallest source of error and does not warrant as much attention.

\section{Summary and Outlook}

In this work, we propose and motivate two new expansions for the dense nuclear matter EoS at finite temperature at arbitrary charge fractions.
We start by demonstrating that a power series expansion in $(T/\mu_B)$ captures the correct thermodynamic scaling near $T=0$ for many physical systems and show that the expansion parameter is small in the regime where the expansion is applied.

We find that the finite $T$ expansion only requires one non-zero term beyond the $T=0$ EoS to reproduce the pressure at $T=100$ MeV at $\mu_B\gtrsim 1150 $ MeV for a fixed $Y_Q$, using an relativistic mean field model as a benchmark.
We break down all the different sources of error in our approach and discuss methods to continue to decrease the error in future work.
Furthermore, our framework provides a clear path for quantifying uncertainty in the finite $T$ EoS relevant for neutron star mergers and outlines a connection to heavy-ion collision experiments through the entropy density over baryon number density expansion. To that end, we highlight the need for experimental data on particle yields, ratios, and fluctuations for charge symmetric ions ($Z/A\approx0.5$), such as $^{16}O$ or $^{24}$Mg, at different $\sqrt{s_{\rm NN}}$. 

The most significant limitation of our approach is the treatment of phase transitions. Near a phase transition, the scaling of thermodynamic variables cannot be captured by a power series in $T/\mu_B$. In the case of first-order phase transitions, there are discontinuities in thermodynamic quantities, and, in the case of second-order phase transitions, these quantities are expected to scale according to a class of critical exponents near the critical point. A possible solution to this problem is to map a parametrization containing the correct behavior near the phase transition to a baseline EoS obtained from the expansion. This type of approach has been implemented successfully using the lattice EoS and a 3D Ising model \cite{Parotto:2018pwx,Karthein:2021nxe,Kapusta:2022pny}.

While our focus in this work has been primarily on neutron star mergers, these techniques could be applied to both heavy-ion collisions and supernova studies.  
For both heavy-ion collisions and neutron star mergers, the uncertainty quantification would need to be studied with more focus on large $Y_Q$ values (for instance, supernovae may reach values of $Y_Q>0.5$ (e.g., \cite{Oertel:2016bki,Kumar:2020gws}), that we did not explore in this work.
For heavy-ion collisions, we would need to match the expanded EoS to the lattice QCD EoS and/or a hadron resonance gas (see, e.g., \cite{Noronha-Hostler:2019ayj,Monnai:2019hkn}) at lower densities. 
Additionally, in heavy-ion collisions, strangeness plays a large role in the EoS and would need to be included in the expansion as well.

At this point, we have only benchmarked our approach using a relativistic mean field EoS containing only proton and neutron degrees of freedom, but we plan to perform future studies with EoS that contain hyperons and/or quarks.
As previously mentioned, our finite $T$ expansion cannot reproduce a first-order phase transition due to the divergence of thermodynamic state variables at the transition, but crossover or higher-order phase transitions into quark or hyperonic phases should be possible if the higher-order terms in the finite $T$ expansion remain small. 

\section*{Acknowledgements}

We thank Jorge Noronha and Mark Alford for useful comments and discussions. We acknowledge support from the support from the US-DOE Nuclear Science Grant No.
DE-SC0023861 and the National Science Foundation under grants PHY1748621, MUSES OAC-2103680, and PHY2309210.  D.M. is supported by the National Science Foundation Graduate Research Fellowship Program under Grant No. DGE-1746047 and the Illinois Center for Advanced Studies of the Universe Graduate Fellowship. We also acknowledge support from the Illinois Campus Cluster, a computing resource that is operated by the Illinois Campus Cluster Program (ICCP) in conjunction with the National Center for Supercomputing Applications (NCSA), which is supported by funds from the University of Illinois at Urbana-Champaign. VD acknowledges additional support from the Department of Energy under grant DE-SC0024700 and from the National Science Foundation under grant NP3M PHY2116686. A.H.\, and L.B.\, are partly supported by the U.S. Department of
Energy, Office of Science, Office of Nuclear Physics, under
Award No. \#DE-FG02-05ER41375. A.H. furthermore acknowledges financial support by
the UKRI under the Horizon Europe Guarantee project
EP/Z000939/1. 

\bibliography{inspire, NOTinspire} 
\onecolumngrid
\appendix

\section{Other thermodynamic quantities for the finite $T$ expansion}\label{app:thermoseries}

It is possible to derive the other thermodynamic quantities directly from Eq.\ (\ref{eq:exp_1}).  
In Eq.\ (\ref{eqn:exp_1}) we consider terms only up to $\mathcal{O}(T^3)$ in the pressure, which is then reflected in our equations below such that we drop high-order terms.
Beginning with the entropy: 
\begin{eqnarray}
    s(T,\vec{\mu})&=& \frac{dp(T,\vec{\mu})}{dT}\biggr\rvert_{\vec{\mu}}\nonumber\\
    &=&2\frac{\partial s(T,\vec{\mu})}{\partial T}\biggr\rvert_{T=0,\vec{\mu}} T+\frac{\partial s^2(T,\vec{\mu})}{\partial T^2}\biggr\rvert_{T=0,\vec{\mu}} T^2+\frac{1}{3}\frac{\partial^3 s(T,\vec{\mu})}{\partial T^3}\biggr\rvert_{T=0,\vec{\mu}} T^3+\mathcal{O}(T^4).
\end{eqnarray}

The same can be done for $n_B$:

\begin{eqnarray}\label{eqn:nBoT3}
    n_B(T,\vec{\mu})&=&\frac{\partial p(T, \vec{\mu})}{\partial \mu _B}\biggr\rvert_{T, \mu_Q} \nonumber\\
    &=& \frac{\partial p(\vec{\mu})}{\partial \mu _B}\biggr\rvert_{T=0, \mu_Q}+ \underbrace{\frac{\partial s(T,\vec{\mu})}{ \partial \mu_B}\biggr\rvert_{T=0, \mu_Q}}_{=0} T + \frac{1}{2}\frac{\partial^2 s(T,\vec{\mu})}{\partial T \partial \mu_B}\biggr\rvert_{T=0, \mu_Q} T^2 +\frac{1}{6}\frac{\partial^3 s(T,\vec{\mu})}{\partial T^2 \partial \mu_B}\biggr\rvert_{T=0, \mu_Q} T^3 +\mathcal{O}(T^4)\nonumber\\
    &=& n_B ( \vec{\mu}) \biggr\rvert_{T=0}+\frac{1}{2}\frac{\partial^2 s(T,\vec{\mu})}{\partial T \partial \mu_B}\biggr\rvert_{T=0, \mu_Q} T^2 +\frac{1}{6}\frac{\partial^3 s(T,\vec{\mu})}{\partial^2 T \partial \mu_B}\biggr\rvert_{T=0, \mu_Q} T^3+\mathcal{O}(T^4) ,
\end{eqnarray}
where $\partial s/\partial \mu_B$ is zero in the limit of $T=0$ since the entropy is zero.

And also for $n_Q$:
\begin{eqnarray}
    n_Q(T,\vec{\mu})&=&\frac{\partial p(T, \vec{\mu})}{\partial \mu _Q}\biggr\rvert_{T, \mu_B} \nonumber\\
    &=& n_Q (\vec{\mu}) \biggr\rvert_{T=0}+\frac{1}{2}\frac{\partial^2 s(T,\vec{\mu})}{\partial T \partial \mu_Q}\biggr\rvert_{T=0, \mu_B} T^2 +\frac{1}{6}\frac{\partial^3 s(T,\vec{\mu})}{\partial T^2 \partial \mu_Q}\biggr\rvert_{T=0, \mu_B} T^3+\mathcal{O}(T^4) .
\end{eqnarray}

We can also calculate higher-order susceptibilities:
\begin{eqnarray}
    \chi_X^n(T,\vec{\mu})&=& \frac{\partial^n p(T, \vec{\mu})}{(\partial \mu_X)^n}\biggr\rvert_{T, \mu_{Y\neq X}} \nonumber\\
    &=& \chi_X^n (\vec{\mu}) \bigr\rvert_{T=0}+\frac{1}{2}\frac{\partial^{n+1} s(T,\vec{\mu})}{\partial T (\partial \mu_X)^n}\biggr\rvert_{T=0, \mu_{Y\neq X}} T^2 +\frac{1}{6}\frac{\partial^{n+2} s(T,\vec{\mu})}{\partial T^2 (\partial \mu_X)^n}\biggr\rvert_{T=0, \mu_{Y\neq X}} T^3+\mathcal{O}(T^4) ,
\end{eqnarray}
where $X=B,Q$.

For a number of EOS it may be sufficient to only expand the pressure up to $\mathcal{O}(T^2)$ and in that case we summarize the corresponding equations:
\begin{eqnarray}
p(T, \vec{\mu})&=&p_{T=0}(\vec{\mu})+\frac{1}{2}\frac{\partial s(T, \vec{\mu})}{\partial T}\biggr\rvert_{T=0,\vec{\mu}} T^2+\mathcal{O}(T^3),\\
    s(T,\vec{\mu})&=&2\frac{\partial s(T,\vec{\mu})}{\partial T}\biggr\rvert_{T=0,\vec{\mu}} T+\frac{\partial^2 s(T,\vec{\mu})}{\partial^2 T}\biggr\rvert_{T=0,\vec{\mu}} T^2+\mathcal{O}(T^3),\\
    n_B(T,\vec{\mu})&=& n_B ( \vec{\mu}) \rvert_{T=0}+\frac{1}{2}\frac{\partial^2 s(T,\vec{\mu})}{\partial T \partial \mu_B}\biggr\rvert_{T=0, \mu_Q} T^2 +\mathcal{O}(T^3)\label{eqn:nBexpan},\\
    n_Q(T,\vec{\mu})&=& n_Q( \vec{\mu}) \rvert_{T=0}+\frac{1}{2}\frac{\partial^2 s(T,\vec{\mu})}{\partial T \partial \mu_Q}\biggr\rvert_{T=0, \mu_B} T^2 +\mathcal{O}(T^3).
\end{eqnarray}

The number of required coefficients are listed for $\mathcal{O}(T^2)$ and $\mathcal{O}(T^3)$ in Table.\ \ref{tab:finiteT_list_coefficients}.  We find that for the finite $T$ expansion, we require only $4$ coefficients for up to $\mathcal{O}(T^2)$ and $7$ coefficients up to $\mathcal{O}(T^3)$.

\begin{table}[h!]
\centering
\begin{tabular}{ | m{10em} | m{2.5cm}| m{2.5cm} | } 
\hline
Thermodynamics & $\mathcal{O}(T^2)$ & $\mathcal{O}(T^3)$\\
\hline
 Pressure &  $\frac{\partial s(T,\vec{\mu})}{\partial T}\bigr\rvert_{T=0,\vec{\mu}}$   & $\frac{\partial s(T,\vec{\mu})}{\partial T}\bigr\rvert_{T=0,\vec{\mu}}$  \\
 & & $\frac{\partial^2 s(T,\vec{\mu})}{\partial^2 T}\bigr\rvert_{T=0,\vec{\mu}}$ \\
 Entropy  & $\frac{\partial s(T,\vec{\mu})}{\partial T}\bigr\rvert_{T=0,\vec{\mu}}$   &  $\frac{\partial s(T,\vec{\mu})}{\partial T}\bigr\rvert_{T=0,\vec{\mu}}$ \\
 & $\frac{\partial^2 s(T,\vec{\mu})}{\partial^2 T}\bigr\rvert_{T=0,\vec{\mu}}$ & $\frac{\partial^2 s(T,\vec{\mu})}{\partial^2 T}\bigr\rvert_{T=0,\vec{\mu}}$ \\
  & & $\frac{\partial^3 s(T,\vec{\mu})}{\partial^3 T}\bigr\rvert_{T=0,\vec{\mu}}$\\
  Baryon Density & $\frac{\partial^2 s(T,\vec{\mu})}{\partial T \partial \mu_B}\bigr\rvert_{T=0, \mu_Q}$&  $\frac{\partial^2 s(T,\vec{\mu})}{\partial T \partial \mu_B}\bigr\rvert_{T=0, \mu_Q}$\\
  &  & $\frac{\partial^3 s(T,\vec{\mu})}{\partial T^2 \partial \mu_B}\bigr\rvert_{T=0, \mu_Q}$ \\
  Charge Density & $\frac{\partial^2 s(T,\vec{\mu})}{\partial T \partial \mu_Q}\bigr\rvert_{T=0, \mu_B}$&  $\frac{\partial^2 s(T,\vec{\mu})}{\partial T \partial \mu_Q}\bigr\rvert_{T=0, \mu_B}$\\
  &  & $\frac{\partial^3 s(T,\vec{\mu})}{\partial T^2 \partial \mu_Q}\bigr\rvert_{T=0, \mu_B}$ \\
  \hline
  Total Unique Terms & 4 & 7 \\
 \hline
\end{tabular}
\caption{List of all needed thermodynamic coefficients  up to orders $T^2$ and $T^3$ for the analytical finite $T$ expansion.}
\label{tab:finiteT_list_coefficients}
\end{table}
\section{Converting between different thermodynamical bases}\label{app:Jacobian}

We want to express $\dfrac{\partial s}{\partial T}\Big | _{\vec{\mu}}$ through the thermodynamical basis $(T,n_B,Y_Q)$. In this basis, the differential of the entropy is 
\begin{equation}
    ds = \dfrac{\partial s}{\partial T}\Big | _{n_B,Y_Q}dT + \dfrac{\partial s}{\partial n_B}\Big | _{T,Y_Q}dn_B + \dfrac{\partial s}{\partial Y_Q}\Big | _{n_B,T}dY_Q,
\end{equation}
so we can calculate
\begin{equation}\label{eqn:APP_dsdTmu}
\dfrac{\partial s}{\partial T}\Big| _{\vec{\mu}} = \dfrac{\partial s}{\partial T}\Big | _{n_B,Y_Q} + \dfrac{\partial s}{\partial n_B}\Big| _{T,Y_Q}\dfrac{\partial n_B}{\partial T}\Big| _{\vec{\mu}} + \dfrac{\partial s}{\partial Y_Q}\Big | _{n_B,T}\dfrac{\partial Y_Q}{\partial T}\Big| _{\vec{\mu}}.
\end{equation} 
We  can then write Eq.\ (\ref{eqn:APP_dsdTmu}) concisely in terms of a matrix product, 
\begin{equation}\label{eq:jacobianmatrix}
 \dfrac{\partial s}{\partial T}\Big| _{\vec{\mu}} = \dfrac{\partial s}{\partial T}\Big | _{n_B,Y_Q} + 
 \begin{pmatrix}
 \dfrac{\partial s}{\partial n_B}\Big| _{T,Y_Q}& 0\\
 0 & \dfrac{\partial s}{\partial Y_Q}\Big | _{n_B,T}\\
 \end{pmatrix}
 \begin{pmatrix}
    \dfrac{\partial n_B}{\partial T}\Big| _{\vec{\mu}} \\ 
    \dfrac{\partial Y_Q}{\partial T}\Big| _{\vec{\mu}} 
 \end{pmatrix}.
\end{equation}

The elements of the column vector specify trajectories of constant $\vec{\mu}$ on the $(n_B,Y_Q)$ plane. These trajectories are given by the constraint equations,
\begin{align}
    0 & = d\mu_B = \dfrac{\partial\mu_B}{\partial T}\Big|_{n_B,Y_Q}dT + \dfrac{\partial\mu_B}{\partial n_B}\Big|_{T,Y_Q}dn_B + \dfrac{\partial\mu_B}{\partial Y_Q}\Big|_{T,n_B}dY_Q,\\
    0 & = d\mu_Q = \dfrac{\partial\mu_Q}{\partial T}\Big|_{n_B,Y_Q}dT + \dfrac{\partial\mu_Q}{\partial n_B}\Big|_{T,Y_Q}dn_B + \dfrac{\partial\mu_Q}{\partial Y_Q}\Big|_{T,n_B}dY_Q,
\end{align}
implying that
\begin{align}\label{eq:constraintseq}
  0 & =  \dfrac{\partial\mu_B}{\partial T}\Big|_{n_B,Y_Q} + \dfrac{\partial\mu_B}{\partial n_B}\Big|_{T,Y_Q}\dfrac{\partial n_B}{\partial T}\Big|_{\vec{\mu}} + \dfrac{\partial\mu_B}{\partial Y_Q}\Big|_{T,n_B}\dfrac{\partial Y_Q}{\partial T}\Big|_{\vec{\mu}},\\
    0 &  = \dfrac{\partial\mu_Q}{\partial T}\Big|_{n_B,Y_Q} + \dfrac{\partial\mu_Q}{\partial n_B}\Big|_{T,Y_Q}\dfrac{\partial n_B}{\partial T}\Big|_{\vec{\mu}} + \dfrac{\partial\mu_Q}{\partial Y_Q}\Big|_{T,n_B}\dfrac{\partial Y_Q}{\partial T}\Big|_{\vec{\mu}}.
\end{align}

Again, we write Eq.~\ref{eq:constraintseq} as a matrix product,
\begin{equation}
-\begin{pmatrix}
    \dfrac{\partial\mu_B}{\partial T}\Big|_{n_B,Y_Q}\\
    \dfrac{\partial\mu_Q}{\partial T}\Big|_{n_B,Y_Q}
\end{pmatrix} = 
\begin{pmatrix}
    \dfrac{\partial\mu_B}{\partial n_B}\Big|_{T,Y_Q} & \dfrac{\partial\mu_B}{\partial Y_Q}\Big|_{T,n_B} \\
    \dfrac{\partial\mu_Q}{\partial T}\Big|_{n_B,Y_Q} & \dfrac{\partial\mu_Q}{\partial Y_Q}\Big|_{T,n_B}
\end{pmatrix}
\begin{pmatrix}
    \dfrac{\partial n_B}{\partial T}\Big|_{\vec{\mu}} \\ 
    \dfrac{\partial Y_Q}{\partial T}\Big|_{\vec{\mu}}
\end{pmatrix}.
\end{equation}

We can now invert the matrix to obtain
\begin{equation}\label{eq:constraintsmatrix}
    \begin{pmatrix}
    \dfrac{\partial n_B}{\partial T}\Big|_{\vec{\mu}} \\ 
    \dfrac{\partial Y_Q}{\partial T}\Big|_{\vec{\mu}}
\end{pmatrix} = -\begin{pmatrix}
    \dfrac{\partial\mu_B}{\partial n_B}\Big|_{T,Y_Q} & \dfrac{\partial\mu_B}{\partial Y_Q}\Big|_{T,n_B} \\
    \dfrac{\partial\mu_Q}{\partial T}\Big|_{n_B,Y_Q} & \dfrac{\partial\mu_Q}{\partial Y_Q}\Big|_{T,n_B}
\end{pmatrix}^{-1}\begin{pmatrix}
    \dfrac{\partial\mu_B}{\partial T}\Big|_{n_B,Y_Q}\\
    \dfrac{\partial\mu_Q}{\partial T}\Big|_{n_B,Y_Q}
\end{pmatrix}.
\end{equation}

Finally, we substitute Eq.~\ref{eq:constraintsmatrix} into Eq.~\ref{eq:jacobianmatrix} to obtain 

\begin{equation}
    \dfrac{\partial s}{\partial T}\Big| _{\vec{\mu}} = \dfrac{\partial s}{\partial T}\Big | _{n_B,Y_Q} - 
 \begin{pmatrix}
 \dfrac{\partial s}{\partial n_B}\Big| _{T,Y_Q}& 0\\
 0 & \dfrac{\partial s}{\partial Y_Q}\Big | _{n_B,T}\\
 \end{pmatrix}
 \begin{pmatrix}
    \dfrac{\partial\mu_B}{\partial n_B}\Big|_{T,Y_Q} & \dfrac{\partial\mu_B}{\partial Y_Q}\Big|_{T,n_B} \\
    \dfrac{\partial\mu_Q}{\partial T}\Big|_{n_B,Y_Q} & \dfrac{\partial\mu_Q}{\partial Y_Q}\Big|_{T,n_B}
\end{pmatrix}^{-1}\begin{pmatrix}
    \dfrac{\partial\mu_B}{\partial T}\Big|_{n_B,Y_Q}\\
    \dfrac{\partial\mu_Q}{\partial T}\Big|_{n_B,Y_Q}
    \end{pmatrix} .
\end{equation}

If we now evaluate this expression at $T=0$,
\begin{equation}
    \dfrac{\partial s}{\partial T}\Big| _{\mu_X,T=0} = \dfrac{\partial s}{\partial T}\Big | _{n_B,Y_Q,T=0} - 
 \begin{pmatrix}
 \dfrac{\partial s}{\partial n_B}\Big| _{T=0,Y_Q}& 0\\
 0 & \dfrac{\partial s}{\partial Y_Q}\Big | _{n_B,T=0}\\
 \end{pmatrix}
 \begin{pmatrix}
    \dfrac{\partial\mu_B}{\partial n_B}\Big|_{T=0,Y_Q} & \dfrac{\partial\mu_B}{\partial Y_Q}\Big|_{T=0,n_B} \\
    \dfrac{\partial\mu_Q}{\partial T}\Big|_{n_B,Y_Q,T=0} & \dfrac{\partial\mu_Q}{\partial Y_Q}\Big|_{T=0,n_B}
\end{pmatrix}^{-1}\begin{pmatrix}
    \dfrac{\partial\mu_B}{\partial T}\Big|_{n_B,Y_Q,T=0}\\
    \dfrac{\partial\mu_Q}{\partial T}\Big|_{n_B,Y_Q,T=0}
    \end{pmatrix},
\end{equation}
knowing that for a system with a non-degenerate ground state  for all $\mu_B$ and $\mu_Q$ we expect $s=0$ at $T=0$, the terms $\dfrac{\partial s}{\partial n_B}\Big| _{T=0,Y_Q}$ and $\dfrac{\partial s}{\partial Y_Q}\Big | _{n_B,T=0}$ should vanish, and thus $\dfrac{\partial s}{\partial T}\Big| _{\mu_X,T=0} = \dfrac{\partial s}{\partial T}\Big | _{n_B,Y_Q,T=0}$ should hold in the $T=0$ limit.
\end{document}